\documentclass[prc,showpacs,showkeys,superscriptaddress,nofootinbib,floatfix,onecolumn]{revtex4}
\usepackage[cp1251]{inputenc}
\usepackage[T2A]{fontenc}
\usepackage[english]{babel}
\usepackage{color}
\usepackage{amsfonts}
\usepackage{amsbsy}
\usepackage{amssymb}
\usepackage{mathrsfs}
\usepackage{graphicx}
\usepackage{subfigure}

\begin{document}
%
\title{Peculiarities of  momentum distribution functions of strongly correlated charged fermions}

\author{A.S.~Larkin$^{1,2}$, V.S.~Filinov$^1$, V.E.~Fortov}

\address{Joint Institute for High Temperatures of the Russian Academy of Sciences, Izhorskaya 13 Bldg 2, Moscow 125412, Russia\\
	$^2$Moscow Institute for Physics and  Technology, Institutskiy per. 9, Dolgoprudny, Moscow Region, 141701,Russia }

\begin{abstract}
The new numerical version of Wigner approach to quantum 
thermodynamics  of strongly coupled 
systems of particles has been developed for extreme conditions, when
analytical approximations obtained in different kind of perturbation
theories can not be applied. Explicit analytical expression of 
Wigner function has been obtained in linear and harmonic
approximations. Fermi statistical effects are accounted by effective
pair pseudopotential depending on coordinates, momenta and
degeneracy parameter of particles and taking into account Pauli
blocking of fermions. The new quantum Monte-Carlo method for
calculations of average values of arbitrary quantum operators has
been proposed. Calculations of the momentum distribution functions and 
pair correlation functions of 
the degenerate ideal Fermi gas have been carried out for testing the
developed approach.
Comparison of obtained momentum distribution functions of strongly
correlated Coulomb systems with Maxwell --  Boltzmann 
and Fermi distributions  shows the significant influence of
interparticle interaction both at small momenta and
in high energy quantum  'tails'.
\end{abstract}



\maketitle

\section{Introduction}\label{Intr} 

Computer simulations are ones of the main tools in quantum statistics nowadays. 
Some of the most powerful numerical methods for simulation of quantum systems are Monte 
Carlo methods, based on path integral (PIMC) formulation of
quantum mechanics \cite{feynman-hibbs}. These methods use path integral representation for 
partition function and thermodynamic values such as average energy, pressure,
heat capacity etc.  Possibility of explicit representation of the partition
function and density matrix in the form of the Wiener path integrals 
\cite{feynman-hibbs,Wiener} as well as making use of Monte Carlo methods
for further calculations allows quantum treatment of strongly coupled systems 
of particles under extreme conditions,
when analytical methods based on perturbation theories can not be applied. 
The main disadvantage of this approach is that PIMC methods cannot cope with 
problem of calculation of average values of arbitrary quantum operators in phase space and  
momentum distribution functions, while this problem may be central in treatment of thermodynamic and 
kinetic properties of matter. 

Considered in this paper Wigner formulation of quantum mechanics in phase space allows 
consideration not only thermodynamic but also   kinetic problems. Quantum effects can affect 
the shape of momentum Maxwellian distribution functions 
since the interaction of a particle with its surroundings restricts the volume of available configuration space, 
which, due to the uncertainty relation, results in an increase in the volume of the momentum space. This results
in increasing fraction of particles with higher momenta 
\cite{Galitskii,Kimball,StarPhys,Eletskii,Emelianov,Kochetov,StMax}. 
Quantum effects are important in studies of such 
phenomena as the transition of combustion into detonation, flame propagation, vibrational relaxation and even 
thermonuclear fusion at high pressure and low temperatures. 
Quantum effects are important in studies of kinetic properties 
of matter at low temperatures and under extreme conditions, when particles are
strongly coupled and perturbative methods cannot be applied. 
 
The main difficulty for path integral Monte Carlo studies of Fermi systems results
from the requirement of antisymmetrization of the density matrix
\cite{feynman-hibbs}. As a result all thermodynamic quantities are
presented as the sum of terms with alternating sign
related to even and odd permutations and are equal to the small
difference of two large numbers, which are the sums of positive and
negative terms. The numerical calculation in this case is severely
hampered. This difficulty is known in the literature as the 'sign
problem'.

To overcome this issue a lot of approaches have been developed. 
Let us mention new original approaches developed in \cite{cpp,prl1,njp,prl2}. 
The configuration path integral Monte Carlo (CPIMC) approach \cite{cpp,prl1}
for degenerate correlated fermions with arbitrary pair interactions
at finite temperatures is based on representation of the
N-particle density operator in a basis of (anti-)symmetrized
N-particle states.
The main idea of this approach is to evaluate the path integral
in space of occupation numbers instead of configuration space
(like in \cite{feynman-hibbs}). This leads to path integrals occupation
number representation allowing to treat arbitrary pair interactions in 
a continuous space. However it turns out that 
CPIMC method exhibits a complementary behavior and works well at 
weak nonideality and strong degeneracy. Unfortunately, the
physically most interesting region, where both fermionic exchange
and interactions are strong simultaneously remains out of reach.

Monte Carlo simulations at finite temperature over the entire fermion density
range down to half the Fermi temperature have been carried out by permutation
blocking path integral Monte Carlo (PB-PIMC) approach \cite{njp,prl2}.
For purpose to simulate fermions in the canonical ensemble, it was
combined a fourth-order approximation of density matrix derived with
a full antisymmetrization on all time
slices in discrete versions of the paths. It was
demonstrated that this approach effectively allows for the combination
of N! configurations from usual PIMC into
a single configuration weight of PB-PIMC, thereby reducing the complexity of
the problem.
Treatment of interacting fermions 
has been carried out at very high densities. Obtained results for finite number of 
particles were extrapolated to the thermodynamic limit. 

Contrary to PIMC methods in configuration space we are going to develop the new numerical approach based on Wigner 
formulation of quantum mechanics \cite{Wigner,Tatar} for treatment of thermodynamic 
properties of non ideal systems of particles in phase space and allowing partially to overcome 
'sign' problem. This new approach allows to analyze  the influence of strong interparticle
interaction on the momentum distribution functions under
extreme conditions, when there are  no small physical parameters and
analytical approximations obtained 
in different kind of  perturbation theories can not be applied. 
Here the new path integral representation of the Wigner function
in the phase space has been developed for canonical ensemble.
Explicit analytical expression of the Wigner function has been
obtained in linear and harmonic approximations. 
Fermi statistical effects are accounted for 
by proposed effective pair pseudopotential. Derived pseudopotential depends on 
coordinates, momenta and degeneracy parameter of fermions and takes into account Pauli blocking 
of fermions in phase space. We have developed 
new quantum Monte-Carlo method for calculations of average values of
arbitrary quantum operators depending on momenta and coordinates.
To test the developed approach calculations of the momentum distribution functions 
and pair  correlation functions of the degenerate ideal system of Fermi particles 
has been carried out in a good agreement with analytical Fermi distributions and available 
pair correlation functions. Comparison of 
obtained momentum distribution function of strongly correlated
Coulomb systems of particles with Maxwell --  Boltzmann and
Fermi distributions  shows the significant influence of
interparticle interaction both at small momenta  and in the high
energy quantum 'tails'.

Let us stress that a simple quasiclassical model of the quantum electron gas based
on a quasiclassical dynamics with an effective Hamiltonian was developed 
in \cite{EbSc97}, where the quantum mechanical effects corresponding to the Pauli and 
the Heisenberg principles were modeled by constraints in the Hamiltonian. 
\section{Wigner function for canonical ensemble}
The Wigner function $W(p,x; t)$ being the analogue of the classical
distribution function in phase space has a wide range of 
applications in quantum mechanics.
Average values of arbitrary physical quantities can
be calculated by formulas similarly to classical statistics. The
Wigner function of the multiparticle system in canonical ensemble
is defined as Fourier transform of the off -- diagonal
matrix element of density matrix in coordinate representation:
\begin{eqnarray}\label{pathint_wignerfunction}
    W(p,x;\beta) = Z(\beta)^{-1}\int{d^{3\tilde{N}}\xi}
    e^{i\langle p|\xi\rangle /\hbar}
    \langle x - \xi/2|e^{-\beta\hat H}|x + \xi/2\rangle ,
\end{eqnarray}
where
$\tilde{N}$ is full number of particle in system, angle brackets in $\langle
p|\xi\rangle$ mean the scalar product of coordinates and momenta,
$\hbar$ is Planck constant and $\beta=1/k_BT$ is proportional to
reciprocal temperature $1/T$. Here we are going to obtain new
representation of Wigner functions in the path integral form
\cite{feynman-hibbs,Wigner,LarkinFilinovCPP,JAMP,Wiener,NormanZamalin,Zamalin},
which allows the numerical simulations of strongly coupled quantum
systems of particles in canonical ensemble.

Since operators of kinetic and potential energy in Hamiltonian do
not commutate, the exact explicit analytical expression for Wigner
function does not in general exist. To overcome this difficulty let
us represent Wigner function in form of path integral
similarly to path integral representation of the partition function
\cite{feynman-hibbs,LarkinFilinovCPP,JAMP,Wiener,NormanZamalin,Zamalin}.
As example of Coulomb system of particles, we consider a 3D  two-component
mass asymmetric electron -- hole plasma consisting of
$N_e$ electrons and $N_h$ heavier holes in equilibrium ($N_e=N_h=N$) \cite{ElHol}. 
The Hamiltonian of the system ${\hat H}={\hat K}+{\hat U}^c$ 
contains kinetic energy ${\hat K}$ and Coulomb interaction energy
${\hat U}^c = {\hat U}_{hh}^c + {\hat  U}^c_{ee} + {\hat
U}^c_{eh}$ contributions. The thermodynamic properties in the
canonical ensemble with given temperature $T$ and fixed volume $V$ are fully
described by the diagonal elements of the density operator
${\hat \rho} = e^{-\beta {\hat H}}/Z$ normalized by the partition function $Z$:
\begin{equation}\label{q-def}
    Z(N_e,N_h,V;\beta) = \frac{1}{N_e!N_h! \lambda_e^{3N_e} \lambda_h^{3N_h} } \sum_{\sigma}\int\limits_V
    dx \,\rho(x, \sigma;\beta),
\end{equation}
where $\rho(x, \sigma;\beta)$ denotes the diagonal matrix
elements of the density operator ${\hat \rho}$.
In equation~(\ref{q-def}), $x=\{x_e,x_h\}$ and $\sigma=\{\sigma_e,\sigma_h\}$
are the spatial coordinates and spin degrees of freedom
of the electrons and holes, i.e.
$x_a=\{x_{1,a}\ldots x_{l,a}\ldots x_{N_a,a}\}$
and $\sigma_a=\{\sigma_{1,a}\ldots \sigma_{t,a}\ldots \sigma_{N_a,a}\}$,
$\lambda_a=\sqrt{\frac{2\pi\hbar^2\beta}{m_a}}$ is the thermal wave length with $a,b=e,h$
and $l,t=1,\dots ,N_a$.

Of course, the exact matrix elements of density matrix of interacting quantum
systems is not known (particularly for low temperatures and high
densities), but they can be constructed using a path integral
approach~\cite{feynman-hibbs,NormanZamalin,Zamalin} based on the operator identity
$e^{-\beta {\hat H}}= e^{-\epsilon {\hat H}}\cdot
e^{-\epsilon {\hat H}}\dots  e^{-\epsilon {\hat H}}$,
where $\epsilon = \beta/M$, which allows us to
rewrite the integral in Eq.~(\ref{q-def}) as
\begin{eqnarray}
&&\sum_{\sigma} \int\limits dx^{(0)}\,
\rho(x^{(0)},\sigma;\beta) =
\int\limits  dx^{(0)} \dots dx^{(m)} \dots
dx^{(M-1)} \, \rho^{(1)}\cdot\rho^{(2)} \, \dots \rho^{(M-1)} \times
\nonumber\\
&&\sum_{\sigma}\sum_{P_e} \sum_{P_h}(\pm 1)^{\kappa_{P_e}+ \kappa_{P_h}} \,
{\cal S}(\sigma, {\hat P_e}{\hat P_h} \sigma_{a}^\prime)\, 
{\hat P_e} {\hat P_h}\rho^{(M)}\big|_{x^{(M)}= x^{(0)}, \sigma'=\sigma}\,.
 \label{rho-pimc}
\end{eqnarray}
The spin gives rise to the spin part of the density matrix (${\cal
S}$) with exchange effects accounted for by the permutation
operators  $\hat P_e$ and $\hat P_h$ acting on the electron and
hole coordinates $x^{(M)}$ and spin projections $\sigma'$. The
sum is taken over all permutations with parity $\kappa_{P_e}$ and
$\kappa_{P_h}$. In Eq.~(\ref{rho-pimc}) the index $m=0,\dots , M-1$
labels the off -- diagonal high-temperature density matrices
$\rho^{(m)}\equiv \rho\left(x^{(m)},x^{(m+1)};\epsilon \right) =
\langle x^{(m)}|e^{-\epsilon {\hat H}}|x^{(m+1)}\rangle$. With the error of order $1/M^2$
arising from neglecting commutator $\epsilon^2/2 \left[K,U^c\right]$ the each high temperature factor can
be presented in the form
$\langle x^{(m)}|e^{-\epsilon {\hat H}}|x^{(m+1)}\rangle \approx
 \langle x^{(m)}|e^{-\epsilon {\hat U^c}}|x^{(m+1)}\rangle \rho^{(m)}_0$,
 where $  \rho^{(m)}_0=\langle x^{(m)}|e^{-\epsilon {\hat K}}|x^{(m+1)}\rangle$.
In the limit $M\rightarrow \infty$ the error of the whole product of high temperature factors is equal to zero $(\propto 1/M)$ 
and we have exact path integral representation of the partition function in which
%
each particle is represented by a trajectory consisting of a set of
$M$ coordinates (``beads''). So the whole configuration of the particles is represented by a
$3(N_e+N_h)M$-dimensional vector
$\tilde{x}\equiv\{x_{1,e}^{(0)}, \dots x_{1,e}^{(M-1)},
x_{2,e}^{(0)}\ldots x_{2,e}^{(M-1)}, \ldots x_{N_e,e}^{(M-1)};
x_{1,h}^{(0)}\ldots x_{N_h,h}^{(M-1)} \}$.
\begin{figure}[htb]
\begin{center}
\includegraphics[width=5cm,clip=true]{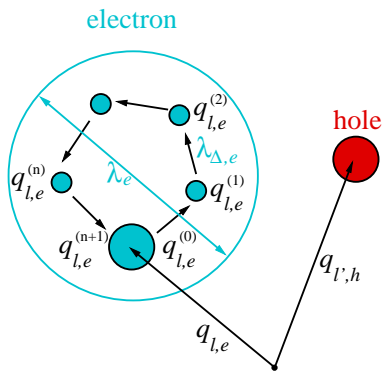}\\
\caption{(Color online) Beads representation of
electrons and holes. Here $\lambda_e^2=2\pi\hbar^2\beta/m_e$,
$\lambda_{\Delta, e}^2=2\pi\hbar^2\epsilon/m_e$
and $\sigma=\sigma^{\prime}$. The holes have a similar beads representation.
For simplicity bead distribution of holes is not
resolved in the figure.} \label{beads}
\end{center}
\end{figure}
Figure.~\ref{beads} illustrates the representation of one (light)
electron and one (heavy) hole. The circle around the electron
beads symbolizes the region that mainly contributes to the
path integral partition function. The size of this region is of
the order of the thermal electron wavelength $\lambda_e$, while
typical distances  between electron beads are of the order of the
electron wavelength taken at an $M$-times higher temperature.
The same representation is valid for each hole but it is not shown.
In the simulations discussed below
the holes are treated according to the full beads representation. 

The problem of approximation of factoring the exponential of Hamiltonian 
$e^{-i \epsilon {\hat H}} \approx e^{-i \epsilon {\hat U^c}} e^{-i \epsilon {\hat K}}$   
in time-dependent quantum-mechanical methods has been considered in details in 
\cite{Kosloff}.  For time-dependent problems short-time propagators are
described: the second-order differencing and the split operator. The time-independent 
approaches based on a polynomial expansion of the evolution operator such as 
the Chebychev propagation and the Lanczos recurrence, have been suggested and considered. 
These methods as applied to many particle systems result to additional computational problems 
in comparison with used here simple approximation. Investigation of advantage of methods \cite{Kosloff} 
 require special work. 

\section{Pair approximation in high-temperature asymptotics for density matrix.}  
In this section we discuss approximations for the high-temperature
density matrix useful for
efficient path integral Monte Carlo (PIMC)
simulations. It involves the effective exchange pseudopotential allowing for
Pauli blocking ( $\tilde{v}^a_{lt}$) and pseudopotential of interparticle
interaction ($\Phi_{ab}$).
Here, we closely follow \cite{Huang} and earlier work \cite{FiBoEbFo01}, where details and further references
can be found.
\subsection{Exchange effects in pair approximation}
To explain the basic ideas of our approach it is
enough to consider the system of ideal electrons and holes, so here
$\hat{U}\equiv0$. The hamiltonian of the system (${\hat H}={\hat
K}={\hat K_e}+{\hat K_h}$) consists
of kinetic energy of electrons ${\hat K_e}$ and holes ${\hat K_h}$.
Due to the commutativity of these operators the
representation of density matrix (\ref{rho-pimc}) is exact at any
finite number $M$. For our purpose it is enough to consider the sum
over permutations in pair approximation at $M=1$:
\begin{eqnarray}
    &&\sum_{\sigma}\sum_{P_e} \sum_{P_h}(\pm 1)^{\kappa_{P_e}+ \kappa_{P_h}} \,
    {\cal S}(\sigma, {\hat P_e}{\hat P_h} \sigma_{a}^\prime)\, 
    {\hat P_e} {\hat P_h}\rho \,
    \big|_{x^{1}= x^{(0)}, \sigma'=\sigma} 
    \nonumber\\ &&
    =\sum_{\sigma_e}\sum_{P_e}(\pm 1)^{\kappa_{P_e}}{\cal S}(\sigma_e, {\hat P_e} \sigma_{e}^\prime)\,
    \rho_e \big|_{x_e^{1}= x_e^{(0)}, \sigma_e'=\sigma_e} \,
    \nonumber\\ &&
    \times \sum_{\sigma_h}\sum_{P_h}(\pm 1)^{\kappa_{P_h}}{\cal S}(\sigma_h, {\hat P_h} \sigma_{h}^\prime)\,
    \rho_h \big|_{x_h^{1}= x_h^{(0)}, \sigma_h'=\sigma_h} 
    \nonumber\\ &&
    =\sum_{\sigma_e}\left \{ 1-\sum_{l<t}f^2_{e;lt}+\sum_{l,t,c}f_{e;lt}f_{e;lc}f_{e; tc}-\dots \right \}
    \nonumber\\ &&
    \times \sum_{\sigma_h}\left \{ 1-\sum_{l<t}f^2_{h;lt}+\sum_{l,t,c}f_{h;lt}f_{h;lc}f_{h; tc}-\dots \right \} \approx\,
    \nonumber\\ &&
    \approx\,\sum_{\sigma_e} \prod_{l<t}(1-f^2_{e;lt})\sum_{\sigma_h}\prod_{l<t}(1-f^2_{h;lt}) =
    \sum_{\sigma}\exp (-\beta \sum_{l<t}\tilde{v}^{e}_{ lt})\exp (-\beta \sum_{l<t}\tilde{v}^{h}_{ lt})
    \label{rho-fermi}
\end{eqnarray}
where  
\begin{eqnarray}
	&&f_{a;lt}=\exp (-\frac{\pi|x^{(0)}_{l,a}-x^{(0)}_{t,a}|^2}{\lambda_a^2}) )
	    \nonumber\\ && 
    \tilde{v}^{a}_{ lt}=-kT \ln(1-f_{a;lt}f_{a; tl})=-kT \ln(1-\delta_{\sigma_{l,a}\sigma_{t,a}}
    \exp (-\frac{2\pi|x^{(0)}_{l,a}-x^{(0)}_{t,a}|^2}{\lambda_a^2}))
    \label{pairfermi}
\end{eqnarray}
is exchange potential \cite{Huang}. This formula shows that the
first corrections accounting for the antisymmetrization of the
density matrix result in the endowing particles by the pair exchange
potential $\tilde{v}^a_{ lt}$.  Below to take into account exchange
effects in Wigner functions we are going to use analogous pair
potential depending on the phase space variables. 
\subsection{Kelbg potential} 
High-temperature density matrix $\rho^{(m)}=\langle
x^{(m)}|e^{-\epsilon {\hat H}}|x^{(m+1)}\rangle$ can be expressed in
terms of two-particle density matrices (higher order terms become
negligible at sufficiently high temperature $M/\beta$
\cite{FiBoEbFo01} )
given by
\begin{eqnarray}\label{rho_ab}
\rho_{ab}(x_{l,a},x'_{l,a}, x_{t,b}, x'_{t,b};\beta)
&=& \frac{(m_a m_b)^{3/2}}{(2 \pi \hbar \beta)^3}
\exp\left[-\frac{m_a}{2 \hbar^2 \beta} (x_{l,a} - x'_{l,a})^2\right]
\, \nonumber\\
&&\times \exp\left[-\frac{m_b}{2 \hbar^2 \beta} (x_{t,b} - x'_{t,b})^2\right]
\exp[-\beta \Phi^{OD}_{ab}]\,.
\end{eqnarray}
This results from factorization of density matrix
into kinetic and potential parts,
$\rho_{ab}\approx\rho_0^K\rho^{U}_{ab}$.
The off -- diagonal density matrix element (\ref{rho_ab}) involves
an effective pair interaction which is expressed approximately via
its diagonal elements, $\Phi^{OD}_{ab}(x_{l,a},x'_{l,a},x_{t,b},
x'_{t,b};\beta)\approx [\Phi_{ab}(x_{l,a}-x_{t,b};
\beta)+\Phi_{ab}(x'_{l,a}-x'_{t,b};\beta)]/2$, for which we use the
well known Kelbg potential
\cite{Ke63,kelbg}
\begin{eqnarray}
\Phi_{ab}(x_{lt};\epsilon) =
\frac{e_a e_b}{\lambda_{ab} x_{lt}} \,\left[1-e^{-x_{lt}^2} +
\sqrt{\pi} x_{lt} \left(1-{\rm erf}(x_{lt})\right) \right],
\label{kelbg-d}
\end{eqnarray}
where $e_a$ and $e_b$ are charges of particles,  $x_{lt}=|x_{l,a}-x_{t,b}|/\lambda_{ab}$,
$\lambda_{ab}=\sqrt{\hbar^2 \epsilon /(2 m_{ab})},$
$m_{ab}=m_{a}m_{b}/(m_{a}+m_{b})$ is the reduced mass of the
$(ab)$-pair of particles and the error function is defined by ${\rm
erf}(x)=\frac{2}{\sqrt{\pi}}\int_0^x dt e^{-t^2}$.

Note that the Kelbg potential is finite at zero distance, which 
is a consequence of allowing for quantum
effects. The validity of this potential and its off--diagonal
approximation are restricted to temperatures substantially higher than the binding energy
\cite{afilinov-etal.04pre,ebeling_sccs05} which puts lower bound on
the number of time slices $M$. For discussion of other
effective potentials, we refer to
 \cite{FiBoEbFo01,afilinov-etal.04pre,ebeling_sccs05,KTR94}.

Summarizing the above approximations, we can conclude that
approximations Eqs.(\ref{rho_ab},\ref{kelbg-d}) of each high-temperature 
factors on the r.h.s. of Eq. (\ref{rho-pimc}) carries an
error of the order $1/M^2$. Within these approximations, we obtain
the result
\begin{eqnarray}
\rho^{(m)}=e^{-\epsilon U(x^{(m)},x^{(m+1)})}\rho_0^{(m)}
+{\cal O}[(1/M)^2]
\nonumber
\label{kel}
\end{eqnarray}
where
 $U$ denotes the sum of all interaction energies, each consisting of the
respective sum of pair interactions given by Kelbg potentials,
$U(x^{(m)},x^{(m+1)})=(U_{hh}(x_h^{(m)})+U_{ee}(x_e^{(m)})+U_{eh}(x_h^{(m)},x_e^{(m)})+U_{hh}(x_h^{(m+1)})+U_{ee}(x_e^{(m+1)})+U_{eh}(x_h^{(m+1)},x_e^{(m+1)}))/2$.
Product of high temperature matrix elements $\rho^{(m)}$ in
Eq.~(\ref{rho-pimc})  has the total error
proportional to $1/M$ and in the limit $M \to
\infty$ presents the exact path integral representation of partition
function \cite{FiBoEbFo01}.

\section{Path integral representation of Wigner function}
Average value of arbitrary quantum operator $\hat A$  can be written
as Weyl's symbol $A(p,x)$ averaged over phase space with the
Wigner function $W(p,x;\beta)$ as weight:
\begin{eqnarray}\label{wigfunc_average0}
\langle \hat A\rangle  = \int\frac{dp dx} {(2\pi\hbar)^{6N}}
A(p,x) W(p,x;\beta),
\end{eqnarray}
where the Weyl's
symbol of operator $\hat A$ is :
\begin{eqnarray}\label{wigfunc_weylsymbol}
A(p,x) = \int d\xi e^{-i\langle
\xi|p\rangle /\hbar} \langle x - \xi/2|\hat A|x + \xi/2\rangle .
\end{eqnarray}
Weyl's symbols for
usual operators like $\hat p$, $\hat x$, $\hat p^2$, $\hat x^2$,
$\hat H$, $\hat H^2$ etc. can be easily calculated directly from
definition (\ref{wigfunc_weylsymbol}).

%

Antisymmetrized Wigner function can be written in the form:
\begin{eqnarray}\label{permut}
&&W(p,x;\beta) = \frac{1}{Z(\beta)N_e!N_h! \lambda_e^{3N_e} \lambda_h^{3N_h} }
\sum_{\sigma}\sum_{P_e} \sum_{P_h}(\pm 1)^{\kappa_{P_e}+
 \kappa_{P_h}} {\cal S}(\sigma, {\hat P_e}{\hat P_h} \sigma^\prime)
\big|_{\sigma'=\sigma}\, \nonumber    \\&& \times \int d \xi
e^{i\langle \xi |p\rangle /\hbar} \langle x - \xi/2|
\prod_{m=0}^{M-1}
 e^{-\epsilon {\hat U_{m}^c}} e^{-\epsilon {\hat K_{m}}}
 |{\hat P_e} {\hat P_h}(x  + \xi/2)\rangle \,
\label{rho_s}
\end{eqnarray}

Now replacing variables of integration $x^{(m)}$ with $q^{(m)}$ for any given permutation $P_eP_h$:
\begin{eqnarray}\label{pathint_variableschange}
&&x_e^{(m)} = (P_e x_e -x_e)\frac{m}{M}+x_e + q^{(m)}_e
-\frac{ (M-m) \xi_e}{2M}+\frac{ m P_e\xi_e }{2M}  \,
\nonumber\\ &&
x_h^{(m)} = (P_h x_h -x_h)\frac{m}{M}+x_h + q^{(m)}_h
-\frac{ (M-m) \xi_h}{2M}+\frac{ m P_h\xi_h }{2M}  \,
\end{eqnarray}
we obtain
\begin{eqnarray}\label{pathint_wignerfunctionint2}
&&W(p,x;\beta) =
\frac{C(M)}{Z(\beta)N_e!N_h! \lambda_e^{3N_e} \lambda_h^{3N_h}  }  \sum_{\sigma}\sum_{P_e} \sum_{P_h}(\pm 1)^{\kappa_{P_e}+
 \kappa_{P_h}} {\cal S}(\sigma, {\hat P_e}{\hat P_h} \sigma^\prime)
\big|_{\sigma'=\sigma}\,
\nonumber    \\&&
 \int d \xi \,
\int   dq^{(1)} \dots dq^{(M-1)}
  \exp\Biggl\{-\pi \frac{\langle \xi|P_eP_h+E |\xi \rangle}
{2M } + i\langle \xi |p\rangle
-\pi \frac{|P_eP_h x-x|^2}{M}
    \nonumber\\&&
    -\sum\limits_{m = 0}^{M-1}
    \biggl[\pi | q^{(m)} - q^{(m+1)}|^2   +
    \epsilon U\biggl((P_eP_h x -x)\frac{m}{M}+x + q^{(m)}
-\frac{ (M-m) \xi}{2M}+\frac{ m P_eP_h\xi}{2M} \biggr)
    \biggr]
    \Biggr\}
    \nonumber \\
\end{eqnarray}
where $E$ is unit matrix, while matrix presenting permutation  $P_eP_h$ is equal to unit 
matrix with appropriately transposed columns. Here and further we
imply that momentum and coordinate are dimensionless variables like
$p_{l,a} \tilde{\lambda}_a/ \hbar$ and $x_{l,a}/ \tilde{\lambda}_a$,
where
$\tilde{\lambda}_a=\sqrt{\frac{2\pi\hbar\beta}{m_a M}}$.
Here constant $C(M)$ as will be shown further is canceled in calculations
of average values of operators.

When $\epsilon \to 0$ this multiple integral turns to exact 
representation of the Wigner function $W(p,x;\beta)$ in the form of
path integral with continuous dimensionless 
'imaginary time' $\tau$ \cite{feynman-hibbs}, which
corresponds to $m/M $ in discrete case. Also the set of independent
variables $q^{(m)}$ turns into closed trajectory $ q(\tau)$.
This trajectory starts and ends in $0$ 
when $\tau = 0$ and $1$.

Let us note that integration here relates to the integration over
the Wiener measure of all closed trajectories $q(\tau)$.
In fact, a particle is presented by the trajectory
with characteristic size of order
$\lambda_a=\sqrt{\frac{2\pi\hbar\beta}{m_a}}$ in coordinate space.
This is manifestation of the uncertainty principle.

\section{Harmonic and linear approximation for Wigner function}
The expression for Wigner function (\ref{pathint_wignerfunctionint2}) is inconvenient
for Monte Carlo simulations because it does not
contain explicit result of integration over $\xi$
even in case of free particle ($U(x) = 0$).
In general this integral can not be calculated
analytically. Exclusions are the linear or harmonic potentials.

To do this integration analytically and to obtain  explicit
expression for Wigner function let us take the approximation for
potential $U(x)$ arising from the Taylor expansion up to the first
or second order in the variables $\xi$:
\begin{eqnarray}\label{harmapprox_potential}
&&U\biggl((P_eP_h x -x)\frac{m}{M}+x + q^{(m)}
-\frac{ (M-m) \xi}{2M}+\frac{ m P_eP_h\xi}{2M} \biggr)
    \approx
U\biggl((P_eP_h x -x)\frac{m}{M}+x + q^{(m)} \biggr)
      \nonumber\\&&
    -\Biggl\langle \frac{ (M-m) \xi}{2M}-\frac{ m P_eP_h\xi}{2M} \bigg|
    \frac{\partial U((P_eP_h x -x)\frac{m}{M}+x + q^{(m)})}{\partial x}
    \Biggr \rangle  +
    \nonumber\\&&
    +\frac{1}{2}\Biggl \langle \frac{ (M-m) \xi}{2M}-\frac{ m P_eP_h\xi}{2M}\bigg|
    \frac{\partial^2 U((P_eP_h x -x)\frac{m}{M}+x + q^{(m)})}{\partial x^2}
    \, \bigg|
    \frac{ (M-m) \xi}{2M}-\frac{ m P_eP_h\xi}{2M} \Biggr\rangle
.
\end{eqnarray}
Here the second term means scalar product of the vector related to
combination of $\xi$ and the  multidimensional gradient of
pseudopotential, while third term means quadratic form with the matrix
of the second derivatives.

This approximation for Wigner function takes the form of gaussian
integral and can be calculated analytically. Here for simplicity let
us consider expressions related to linear approximation accounting
for the linear term in expansion. Accounting for quadratic term is obvious.
Then the Wigner function can be written in the following form:
\begin{eqnarray}
\label{pathint_wignerfunctionint3}
&&W(p,x;\beta) =
\frac{C(M)}{Z(\beta)N_e!N_h! \lambda_e^{3N_e} \lambda_h^{3N_h} }  \sum_{\sigma}\sum_{P_e} \sum_{P_h}(\pm 1)^{\kappa_{P_e}+
 \kappa_{P_h}} {\cal S}(\sigma, {\hat P_e}{\hat P_h} \sigma^\prime)
\big|_{\sigma'=\sigma}\,
\nonumber    \\&&
\times \int   dq^{(1)} \dots dq^{(M-1)}
  \exp\Biggl\{
    -\sum\limits_{m = 0}^{M-1}
    \biggl[\pi | q^{(m)} - q^{(m+1)}|^2   +
    \epsilon U\biggl((P_eP_h x -x)\frac{m}{M}+x + q^{(m)}  \biggr)
    \biggr]
    \Biggr\}
    \nonumber\\&&
\times  \exp\Biggl\{-\pi \frac{|P_eP_h x-x|^2}{M} \Biggr\}  \int d \xi \, \exp\Biggl\{-\pi \frac{\langle \xi|P_eP_h+E |\xi \rangle} {2M }
    \nonumber\\&&
    + \Biggl \langle \xi \Bigg| i p
    + \sum\limits_{m = 0}^{M-1}\bigg|
    \frac{ (M-m) }{2M} E -\frac{ m }{2M} P_eP_h  \bigg|
    \frac{\partial \epsilon U(\bar{x})}{\partial \bar{x}}
    \Biggr \rangle
    \Biggr|_{\bar{x}=(P_eP_h x -x)\frac{m}{M}+x + q^{(m)}}
    \Biggr\}.
\end{eqnarray}
This expression for Wigner function
is obtained under assumption that potential energy $U$ is expandable in Taylor series on
$\xi$ with a good accuracy. Let us discuss the legality of such assumption for one particle in 1D case and
identical permutation.
The Taylor expansion of potential function contains powers of $\xi$ multiplied on 
$\sum\limits_{m = 0}^{M-1}(m/M -1/2)^n\frac{\partial^n U(x+q_m)}{n! \partial x^n}$.
Using mean value theorem, we can roughly get symbolic estimation of these sums as
\begin{eqnarray}\label{harmapprox_inttauvalues}
\sum\limits_{m = 0}^{M-1}(m/M -1/2)^n\frac{\partial^n U(x+q^{(m)})}{n! \partial x^n } \approx 
\int\limits_{0}^{1}{d\tau} \frac{1}{n!}\frac{\partial^n U(x+q^{(\tau)})}{\partial x^n}
\left( \tau-\frac{1}{2} \right)^n \approx 
\frac{1}{n!}\frac{(1 + (-1)^n)}{(n+1)2^{n+1}}
\frac{\partial^n U(x+q^{(0)})}{\partial x^n},
\end{eqnarray}
where $q^{(0)}=q^{(M)}=0$. 
We expect the fast convergence of the Taylor series for potentials 
averaged along the trajectories $x+q^{(m)}$ with sign alternating for odd $n$ 
weight $\left( \tau-\frac{1}{2} \right)^n$. 
Numerical valuee of this integral rapidly decreases as $1/24$, $1/1920$ and $1/322560$ for $n = 2,4,6$ 
and are lesser for odd $n$. 
To check the accuracy of this approximation 
Monte Carlo simulations of the thermodynamic values of particles in different arbitrary 
potentials have been carried out in \cite{LarkinFilinovCPP,JAMP}.  These calculations confirm 
our expectations. 

As was mentioned before it is enough for our purpose (see Eq.(\ref{rho-fermi})) to take
into account pair permutations. In degenerate system average
distance between fermions is less than the thermal wavelength
$\lambda$ and trajectories in path integrals
(\ref{pathint_wignerfunctionint3}) are strongly entangled.
This is the reason that pair permutations can not strongly affect the potential energy
in (\ref{pathint_wignerfunctionint3})
in comparison with the case of identical permutation. 
So for any pair permutation  the potential energy in (\ref{pathint_wignerfunctionint3})
can be presented as energy related to identical permutation (of order $N^2M/2$) 
and small (in comparison with identical permutation) difference (of order $2NM$):
\begin{eqnarray}\label{small}
\sum\limits_{m = 0}^{M-1}\epsilon U\biggl((P_eP_h x -x)\frac{m}{M}+x + q^{(m)}  \biggr) -\sum\limits_{m = 0}^{M-1} \epsilon U\biggl(x + q^{(m)}  \biggr).
\end{eqnarray}
So we can take only the first terms of the perturbation series on this 
relative difference and can obtain the following simplification:
$\sum\limits_{m = 0}^{M-1}\epsilon U\biggl((P_eP_h x -x)\frac{m}{M}+x + q^{(m)}  \biggr)
\approx \sum\limits_{m = 0}^{M-1}\epsilon U\biggl(x + q^{(m)}  \biggr) $.
Rigorous proof can be done within generalization of Mayer expansion technique (so called algebraic approach) on non ideal
systems of particles developed in \cite{Zelener, Ruelle}.
Now all permutations in (\ref{pathint_wignerfunctionint3}) are
acting only on variables $x$ and $\xi$ and can be beared out
of the path integral.

So the Wigner function is determined by path integral over
all closed trajectories and can be presented in the form
\begin{eqnarray}
\label{pathint_wignerfunctionint4}
&&W(p,x;\beta) \approx\,
\frac{C(M)}{Z(\beta)N_e!N_h! \lambda_e^{3N_e} \lambda_h^{3N_h} }
\int   dq^{(1)} \dots dq^{(M-1)}
\nonumber\\&&  \times
 \exp\Biggl\{
    -\sum\limits_{m = 0}^{M-1}
    \biggl[ \pi | q^{(m)} - q^{(m+1)}|^2   +
    \epsilon U\biggl(x + q^{(m)}
) \biggr]   \Biggr\}
     \nonumber\\&&
     \times \exp\Biggl\{\frac{M}{4 \pi}
     \Biggr|i p + \frac{\epsilon}{2}\sum\limits_{m = 0}^{M-1} \frac{ (M-2m) }{M}
     \frac{\partial  U(x+q^{(m)})}{\partial x}
     \Biggr|^2\Biggr\}
     \nonumber\\&&
\times\sum_{\sigma_e}\Biggl\{1-\sum_{l<t}\delta_{\sigma_{l,e}\sigma_{t,e}}
\exp(-2\pi\frac{|x_{l,e}-x_{t,e}|^2}{M})
\delta \biggl(\frac{(\tilde{p}_{l,e}-\tilde{p}_{t,e})\sqrt{M}}{2\pi}
    \biggr)\Biggr\}
     \nonumber\\&&
\times\sum_{\sigma_h}\Biggl\{1-\sum_{l<t}\delta_{\sigma_{l,h}\sigma_{t,h}}
\exp(-2\pi\frac{|x_{l,h}-x_{t,h}|^2}{M})
\delta \biggl(\frac{(\tilde{p}_{l,h}-\tilde{p}_{t,h})\sqrt{M}}{2\pi}
 \biggr)\Biggr\}
\end{eqnarray}
 where
 \begin{eqnarray}
\tilde{p}_{t,a}=p_{t,a} + \frac{\epsilon }{2 }\sum\limits_{m = 0}^{M-1}
    \frac{\partial  U(x+q^{(m)})}{\partial x_{t,a}}
    \nonumber
  \end{eqnarray}

The main idea of deriving  expression  (\ref{pathint_wignerfunctionint4})
can be explained on example of two electrons in $1D$ space. For two
electrons the sum over permutations in
(\ref{pathint_wignerfunctionint3}) consists of two
terms related to identical permutation (matrix $P$ is equal to unit
matrix $E$) and non identical permutation (matrix $P$ is equal to
matrix $E$ with transposed columns). To do integration in
(\ref{pathint_wignerfunctionint3}) over $\xi$ let us analyze
eigenvalues of matrix $P+E$.  For identical
permutation the eigenvalues are equal to each other and are equal to
two, while the eigenvalues of matrix $P+E$ related to non identical
permutation are equal to zero and two. Integration over $\xi$ for
identical permutation is trivial, while for non identical
permutation matrix $P+E$ have to be
presented in the form $P+E=ODO^{-1}$, where $D$ is diagonal matrix
with zero and two as the diagonal elements. Here matrix $O$ and
inverse matrix $O^{-1}$ are given by the formulas:
\[ O = \left|
\begin{array}{cc}
1 & 1  \\
-1 & 1  \\
\end{array} \right|.\]
\[ O^{-1} = \frac{1}{2}\left| \begin{array}{cc}
1 & -1  \\
1 & 1  \\
\end{array}\right|.\]
Replacing variables on integration by relation $|\zeta, \eta>=|O^{-1}|\xi>$ one can
obtain expression analogous (\ref{pathint_wignerfunctionint4}).

To obtain the final expression 
we have to approximate delta-function by the standard Gaussian exponent with
small parameter $\alpha $ :
\begin{eqnarray}
\label{pathint_wignerfunctionint5}
&&W(p,x;\beta) \approx\,
\frac{C(M)}{Z(\beta)N_e!N_h!\lambda_e^{3N_e} \lambda_h^{3N_h} } \int   dq^{(1)} \dots dq^{(M-1)}
    \nonumber\\&& \times
  \exp\Biggl\{
    -\sum\limits_{m = 0}^{M-1}
    \biggl[\pi | q^{(m)} - q^{(m+1)}|^2   +
    \epsilon U(x + q^{(m)}
) \biggr]   \Biggr\}
    \nonumber\\&&
 \times \exp\Biggl\{\frac{M}{4 \pi}
\Biggr| i p + \frac{\epsilon}{2}\sum\limits_{m = 0}^{M-1} \frac{ (M-2m) }{M}
 \frac{\partial  U(x+q^{(m)})}{\partial x}
\Biggr|^2\Biggr\}
\sum_{\sigma}\exp (-\beta \sum_{l<t}v^{e}_{ lt})\exp (-\beta \sum_{l<t}v^{h}_{ lt})
\Biggr|^2\Biggr\}
\nonumber\\
\end{eqnarray}
where
\begin{eqnarray}
v^{a}_{ lt}=-kT \ln\Biggl\{1-\delta_{\sigma_{l,a}\sigma_{t,a}}
\exp \biggl(-\frac{2\pi|x_{l,a}-x_{t,a}|^2}{M}\biggr)
\frac{\sqrt{M}}{2\pi\alpha}
\exp \biggl(-\pi\frac{|(\tilde{p}_{l,a}-\tilde{p}_{t,a})\sqrt{M}|^2}
{(2\pi\alpha)^2}\biggr)\Biggr\}
     \nonumber
\end{eqnarray}
(For functions $v^{a}_{ lt}$  we can imply here analytic continuation
on complex plane.) Note that the expression
(\ref{pathint_wignerfunctionint5}) contains term
related the classical Maxwell distribution
explicitly. The difference is in the other
terms accounting for influence of
 interaction on the momentum distribution function.
 To regularize integration over  momenta 
 in the limit of small $\alpha$
it is necessary to rescale $p$ by factor
$\frac{\sqrt{M}}{2\pi\alpha}$  in (\ref{wigfunc_average0}), so one can use
the simplified version of
effective pair  exchange pseudopotential ($\pi$ is included in small
$\alpha^2$):
\begin{equation}\label{vex}
v^{a}_{ lt}\approx\,-kT \ln\Biggl\{1-\delta_{\sigma_{l,a}\sigma_{t,a}}
\exp \biggl(-\frac{2\pi|x_{l,a}-x_{t,a}|^2}{\lambda^2_a}\biggr)
\exp \biggl(-\frac{|(\tilde{p}_{l,a}-\tilde{p}_{t,a})\lambda_a|^2}
{(2\pi\hbar)^2\alpha^2}\biggr)\Biggr\}
\nonumber
\end{equation}
Here  momenta and coordinates are written in natural units ($\lambda_a^2=\frac{2\pi\hbar^2 \beta}{m_a}$).
\section{Average values of quantum operators}
To calculate average values of quantum operators $\langle \hat A\rangle $ we are going to
use the Monte Carlo method (MC) \cite{Metropolis,Hasting}. For this aim we have to use  path integrals 
 (\ref{pathint_wignerfunctionint5}) in
discrete form. As a result we obtain final
expressions for MC calculations as follows: 
\begin{eqnarray}\label{mmc_averages}
    \langle \hat A\rangle  =
    \frac{\left\langle A(p,x)
    \cdot h(p,x,q^{(1)},\dots,q^{(M-1)})\right\rangle _w}
    {\langle
    h(p,x,q^{(1)},\dots,q^{(M-1)})\rangle_w}.
\end{eqnarray}
Here brackets $\bigl\langle g(p,x,q^{(1)},\dots,q^{(M-1)})\bigr\rangle _w$ denote
averaging of any function $g(p,x,q^{(1)},\dots,q^{(M-1)})$ with positive
weight $w(p,x,q^{(1)},\dots,q^{(M-1)})$:
\begin{eqnarray}\label{mmc_averagesav}
    \langle  g(p,x,q^{(1)},\dots,q^{(M-1)}) \rangle_w  =
    \int{dp dx}\int{dq^{(1)} \dots dq^{(M-1)}}
    g(p,x,q^{(1)},\dots,q^{(M-1)})
    w(p,x,q^{(1)},\dots,q^{(M-1)})
\end{eqnarray}
while
\begin{eqnarray}\label{mmc_averageswfh}
w(p,x,q^{(1)},\dots,q^{(M-1)}) =
     \Biggr|\cos\Biggr\{
     2 \frac{M}{4 \pi}
\Biggl \langle  p \Bigg|
\frac{\epsilon}{2}\sum\limits_{m = 0}^{M-1} \frac{ (M-2m) }{M}
 \frac{\partial  U(x+q^{(m)})}{\partial x}
\Biggr \rangle \Biggr\}\Biggl|
\nonumber\\
    \times \exp\Biggl\{-\frac{M}{4 \pi} \biggl(
    \Bigg| p\Bigg|^2  +
    \Bigg| \frac{\epsilon}{2}\sum\limits_{m = 0}^{M-1} \frac{ (M-2m) }{M}
    \frac{\partial  U(x+q^{(m)})}{\partial x}
    \Biggr|^2 \biggr) \Biggr\}
    \nonumber\\
\times \exp\Biggl\{-\sum\limits_{m = 0}^{M-1} \biggl[
\pi | q^{(m)} - q^{(m+1)}|^2   +
\epsilon U(x + q^{(m)}   ) \biggr]   \Biggr\}
\sum_{\sigma}\exp (-\beta \sum_{l<t}v^{e}_{ lt})\exp (-\beta \sum_{l<t}v^{h}_{ lt}) ,
    \nonumber\\
h(p,x,q^{(1)},\dots,q^{(M-1)}) = sign \Biggr( \cos\Biggr\{
     2 \frac{M}{4 \pi}
\Biggl \langle  p \Bigg| \frac{\epsilon}{2}\sum\limits_{m = 0}^{M-1} \frac{ (M-2m) }{M}
 \frac{\partial  U(x+q^{(m)})}{\partial x}
\Biggr \rangle \Biggr\} \Biggl).
\end{eqnarray}
Note that denominator in (\ref{mmc_averages}) is equal to nominator with
$A(p,x) = 1$, so
 $C(M)$ from (\ref{pathint_wignerfunctionint5})
is canceled. 

Eq.~(\ref{mmc_averages}) can be rewritten in compact form:  
\begin{eqnarray}
	\langle A\rangle_w=\frac{\sum_{i=1}^N A(p_i,x_i) h(\overline{\mathbf{x}}_i)  }
	{\sum_{i=1}^N h(\overline{\mathbf{x}}_i)}.
	\label{MC3}
\end{eqnarray}  
where $h(\overline{\mathbf{x}}_i)$ is arbitrary weight function of the random quantity 
$\overline{\mathbf{x}}_i$ drawn from any distribution $w(\overline{\mathbf{x}})/\overline{Q}$ 
\ ($
\overline{Q} = \int_{\Omega}w(\overline{\mathbf{x}})d\overline{\mathbf{x}}).
$

Metropolis -- Hastings algorithm resides in designing a Markov process (by constructing transition probabilities 
$P(\overline{\mathbf{x}}\rightarrow \overline{\mathbf{x}}')$ ), such that its stationary distribution 
to be equal to $w(\overline{\mathbf{x}})$. The derivation of the algorithm starts with the condition of detailed 
balance: 
\begin{eqnarray}
	w(\overline{\mathbf{x}})P(\overline{\mathbf{x}}\rightarrow \overline{\mathbf{x}}') = w(\overline{\mathbf{x}}')P(\overline{\mathbf{x}}'\rightarrow \overline{\mathbf{x}})
\end{eqnarray}
which can be rewritten as 
\begin{eqnarray}
	\frac{P(\overline{\mathbf{x}}\rightarrow \overline{\mathbf{x}}')}{P(\overline{\mathbf{x}}'\rightarrow \overline{\mathbf{x}})} = \frac{w(\overline{\mathbf{x}}')}{w(\overline{\mathbf{x}})}. 
\end{eqnarray}
The idea is to separate the transition in two sub-steps:   
the proposal and the acceptance-rejection. 
The transition probability can be written as the product: 
$P(\overline{\mathbf{x}}\rightarrow \overline{\mathbf{x}}') = g(\overline{\mathbf{x}}\rightarrow \overline{\mathbf{x}}') A(\overline{\mathbf{x}}\rightarrow \overline{\mathbf{x}}')$. 
The proposal distribution 
$\displaystyle g(\overline{\mathbf{x}}\rightarrow \overline{\mathbf{x}}')$ is 
the conditional probability of proposing a state $\overline{\mathbf{x}}'$ for given $\overline{\mathbf{x}}$, 
and the acceptance distribution $\displaystyle A(x\rightarrow x')$ is 
the conditional probability to accept the proposed state $\overline{\mathbf{x}}'$. 
Inserting this relation in the previous equation, we have 
\begin{eqnarray}
	\frac{A(\overline{\mathbf{x}}\rightarrow \overline{\mathbf{x}}')}{A(\overline{\mathbf{x}}'\rightarrow \overline{\mathbf{x}})} = \frac{w(\overline{\mathbf{x}}')}{w(\overline{\mathbf{x}})}\frac{g(\overline{\mathbf{x}}'\rightarrow \overline{\mathbf{x}})}{g(\overline{\mathbf{x}}\rightarrow \overline{\mathbf{x}}')}
\end{eqnarray}.

Then it is necessary to choose an acceptance that fulfills detailed balance. 
One common choice is the Metropolis's suggestion:
\begin{eqnarray}
	A(\overline{\mathbf{x}}\rightarrow \overline{\mathbf{x}}') = \min\left(1,\frac{w(\overline{\mathbf{x}}')}{w(\overline{\mathbf{x}})}\frac{g(\overline{\mathbf{x}}'\rightarrow \overline{\mathbf{x}})}{g(\overline{\mathbf{x}}\rightarrow \overline{\mathbf{x}}')}\right)
\end{eqnarray}
This means that we always accept when the acceptance is bigger than 1  
and we can accept or reject when the acceptance is smaller than 1. 
As the conditional probability $g$ we can use 
\begin{eqnarray}
g(\overline{\mathbf{x}}\rightarrow \overline{\mathbf{x}}')= \exp\Biggl\{-\frac{M}{4 \pi} \biggl(\Bigg| p'\Bigg|^2  +
\Bigg| \frac{\epsilon}{2}\sum\limits_{m = 0}^{M-1} \frac{ (M-2m) }{M}
\frac{\partial  U(x'+q^{(m)})}{\partial x'} \Biggr|^2 \biggr) \Biggr\} / \\
\exp\Biggl\{-\frac{M}{4 \pi} \biggl(
\Bigg| p\Bigg|^2  +
\Bigg| \frac{\epsilon}{2}\sum\limits_{m = 0}^{M-1} \frac{ (M-2m) }{M}
\frac{\partial  U(x+q^{(m)})}{\partial x} \Biggr|^2 \biggr) \Biggr\}
\end{eqnarray}. 

The Metropolis -- Hastings algorithm can consist in the following:\\
1) Initialization: pick an initial state point $(p,x,q^{(1)},\dots,q^{(M-1)}) = \overline{\mathbf{x}}$ at random; \\
2) randomly pick a state $x'$ according to $\displaystyle g(\overline{\mathbf{x}}\rightarrow \overline{\mathbf{x}}')$; \\
3) accept the state according to the probability $\displaystyle A(\overline{\mathbf{x}}\rightarrow \overline{\mathbf{x}}')$. 
If not accepted, that means that $\overline{\mathbf{x}}' = \overline{\mathbf{x}}$, and so there is no need to update anything. 
Else, the system transits to $\overline{\mathbf{x}}'$; \\
4) go to 2 until many $M$ states were generated; \\
5) save the state $\overline{\mathbf{x}}$, go to $2$.

It is important to notice that it is not clear, in a general problem, which distribution 
$\displaystyle g(\overline{\mathbf{x}}\rightarrow \overline{\mathbf{x}}')$ one should use. It is a free parameter of the method which 
has to be adjusted to the particular problem 'in hand'. 
This is usually done by calculating the acceptance rate, which is the fraction of 
proposed samples that is accepted during the last $\displaystyle N$ samples. 
As has been shown theoretically the ideal acceptance rate 
have to be in interval of $23$ -- $50$ \%. 
The Markov chain is started from an arbitrary initial value $\displaystyle \overline{\mathbf{x}}_0$ and the algorithm 
is run for many iterations until this initial state is "forgotten". These samples, which are 
discarded, are known as 'burn-in'. The remaining set of accepted values of $\overline{\mathbf{x}}$ represent 
a sample from the distribution $w(\overline{\mathbf{x}})$.

To construct Metropolis -- Hasting algorithm  \cite{Metropolis,Hasting} we introduce three types of PIMC  steps:
\begin{enumerate}
    \item {Variation of momentum of some particle.} 
    \item {Variation of overall position of some particle.} 
    \item {Variation of trajectory representing some particle.} 
\end{enumerate}
The first thousands of steps should be rejected during the
calculation to "forget" initial configuration. Periodic boundary
conditions have to be used to reduce finite number effects \cite{Metropolis}. 

Calculations of the average values of quantum operators depending only on coordinates of particles 
is more convenient and reasonable to carry out in configurational space  by standard path integral 
Monte Karlo method (PIMC).  Within considered above approach it can be done  if we change 
the following functions: 
\begin{eqnarray}\label{mc_averageswfh}
	&&\tilde{w}(x,q^{(1)},\dots,q^{(M-1)}) =
	\exp\Biggl\{-\sum\limits_{m = 0}^{M-1} \biggl[
	\pi | q^{(m)} - q^{(m+1)}|^2   +
	\epsilon U(x + q^{(m)}   ) \biggr]   \Biggr\}
	\sum_{\sigma}\exp (-\beta \sum_{l<t}\tilde{v}^{e}_{ lt})\exp (-\beta \sum_{l<t}\tilde{v}^{h}_{ lt}) ,
	\nonumber\\ &&
	\tilde{h}(x,q^{(1)},\dots,q^{(M-1)}) \equiv 1 .
\end{eqnarray}
where $\tilde{v}^{a}_{ lt}$ is defined by equation~(\ref{pairfermi}),  
\begin{eqnarray}\label{mcg_averagesav}
	\langle  g(x,q^{(1)},\dots,q^{(M-1)}) \rangle_{\tilde{w}}=
	\int{dx}\int{dq^{(1)} \dots dq^{(M-1)}}
	g(x,q^{(1)},\dots,q^{(M-1)})
	\tilde{w}(x,q^{(1)},\dots,q^{(M-1)})
\end{eqnarray}
and 
\begin{eqnarray}\label{mm_averages}
	\langle  {\tilde{ \hat A}}\rangle  =
	\frac{\left\langle \tilde{A}(x)
		\cdot \tilde{h}(x,q^{(1)},\dots,q^{(M-1)})\right\rangle_{\tilde{w}}}
	{\langle
		\tilde{h}(x,q^{(1)},\dots,q^{(M-1)})\rangle_{\tilde{w}}}.
\end{eqnarray}
\section{Results of numerical calculations}
We define momentum distribution functions and pair correlation functions  
for holes ($a=h$) and electrons ($a=e$) by the following expressions:
\begin{eqnarray}  \label{gab-rho}
&& w_a(|p|) =  
	    \frac{\left\langle \delta(|p_{1,a}|-|p|) 
	    	\cdot h(p,x,q^{(1)},\dots,q^{(M-1)})\right\rangle _w}
	    {\langle
	    	h(p,x,q^{(1)},\dots,q^{(M-1)})\rangle_w} 
\nonumber\\&&   
	 g_{ab}(r)  =
	\frac{\left\langle \delta(|x_{1,a}-x_{1,b}|-r) 
		\cdot \tilde{h}(x,q^{(1)},\dots,q^{(M-1)})\right\rangle_{\tilde{w}}}
	{\langle
		\tilde{h}(x,q^{(1)},\dots,q^{(M-1)})\rangle_{\tilde{w}}} 
\end{eqnarray}
where $\delta$ is delta function, $a$ and $b$ are types of the particles. The pair
correlation functions $g_{ab}$ give a probability density to find a
pair of particles of types $a$ and $b$ at a certain distance $r$
from each other and depend only on the difference of coordinates
due to the translational
invariance of the system. In a noninteracting classical system,
$g_{ab}  \equiv 1 $, whereas interactions and quantum statistics 
result in a spatial redistribution of the particles.  

The momentum distribution function $w_a(|p|)$ gives
a probability density for particle of type $a$ to have momentum $p$.
Non ideal classical systems of particles due to the commutativity of
kinetic and potential energy operators  have Maxwell distribution
function (MD) proportional to $\exp(-(p\lambda_a)^2/4\pi\hbar^2)$
even at strong coupling. Quantum effects can affect the
shape of kinetic energy distribution function.
Quantum ideal systems of particles due to the
quantum statistics have Fermi or Bose momentum distribution
functions. Interaction of a quantum particle with its
surroundings restricts the volume of configuration space,
which,  can also affect the
shape of momentum distribution function due to the uncertainty relation.
In this section we present results for both ideal and strongly coupled electron -- hole plasmas 
with different hole masses.

\textbf{Test calculations for ideal electron -- hole plasma}. 
To extent the region of applicability of pair approximation the small parameter $\alpha^2$ 
in Eq.~(\ref{vex})  
has been used as adjustable function of the universal degeneracy parameter of ideal fermions
$n\lambda^3$, namely $\alpha^2_a=0.00505+0.056n\lambda_a^3$. 
To simplify calculations we fix the number of
electrons and holes with the same spin projection equal to $N_e/2$
and $N_h/2$ respectively. 
To test the developed approach calculations of the path
integral representation of Wigner function
in the form (\ref{mmc_averages}) have been carried out for different number of particles 
and beads presenting each particles. Results have been obtained by averaging-out over 
several hundred thousands and several millions Monte Carlo steps. 
It turn out that for convergence is enough several hundred thousand Monte Carlo steps and two 
hundred particles each presented by twenty beads. 
\begin{figure}[htb]
    \includegraphics[width=8.9cm,clip=true]{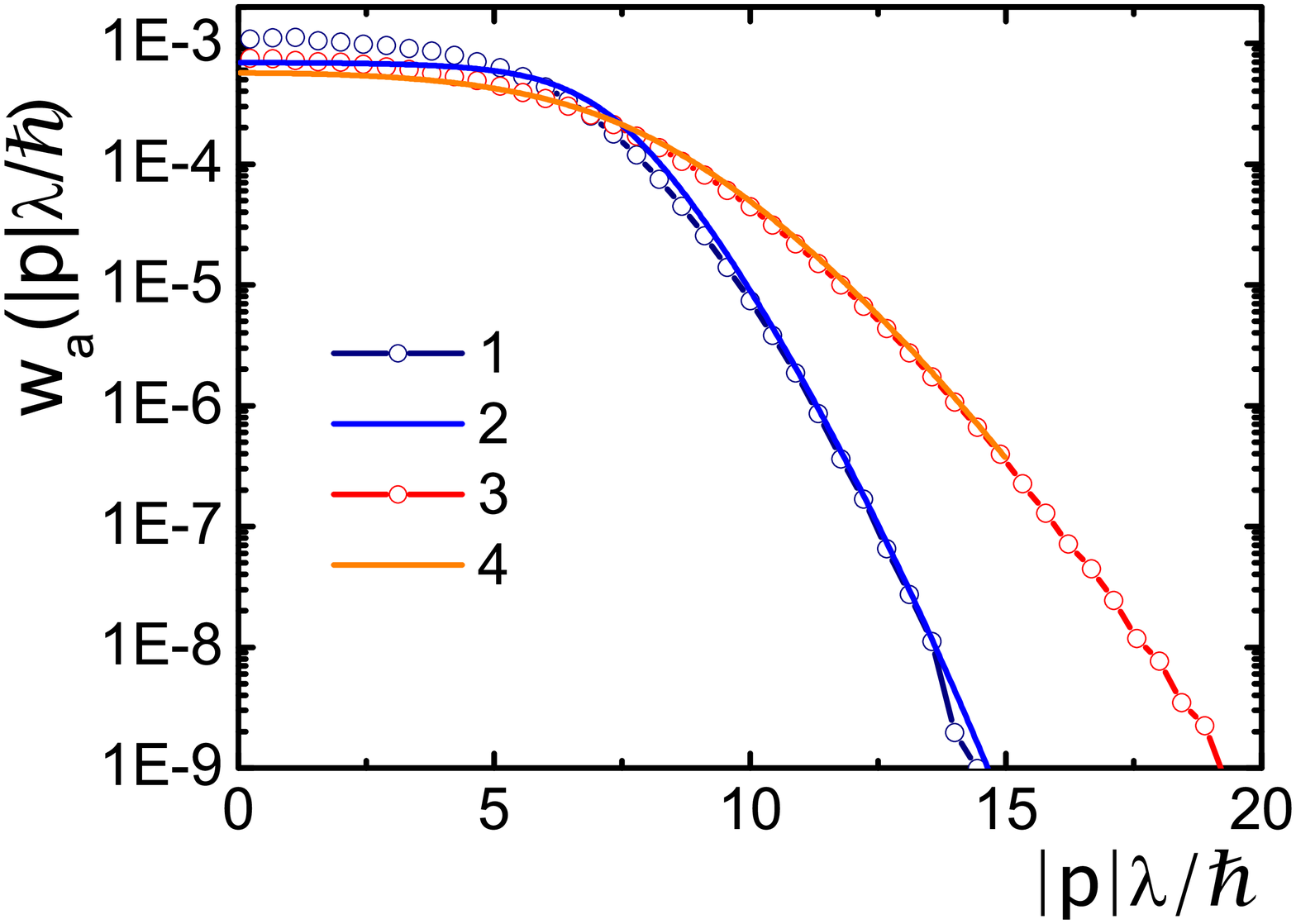}              
    \includegraphics[width=8.9cm,clip=true]{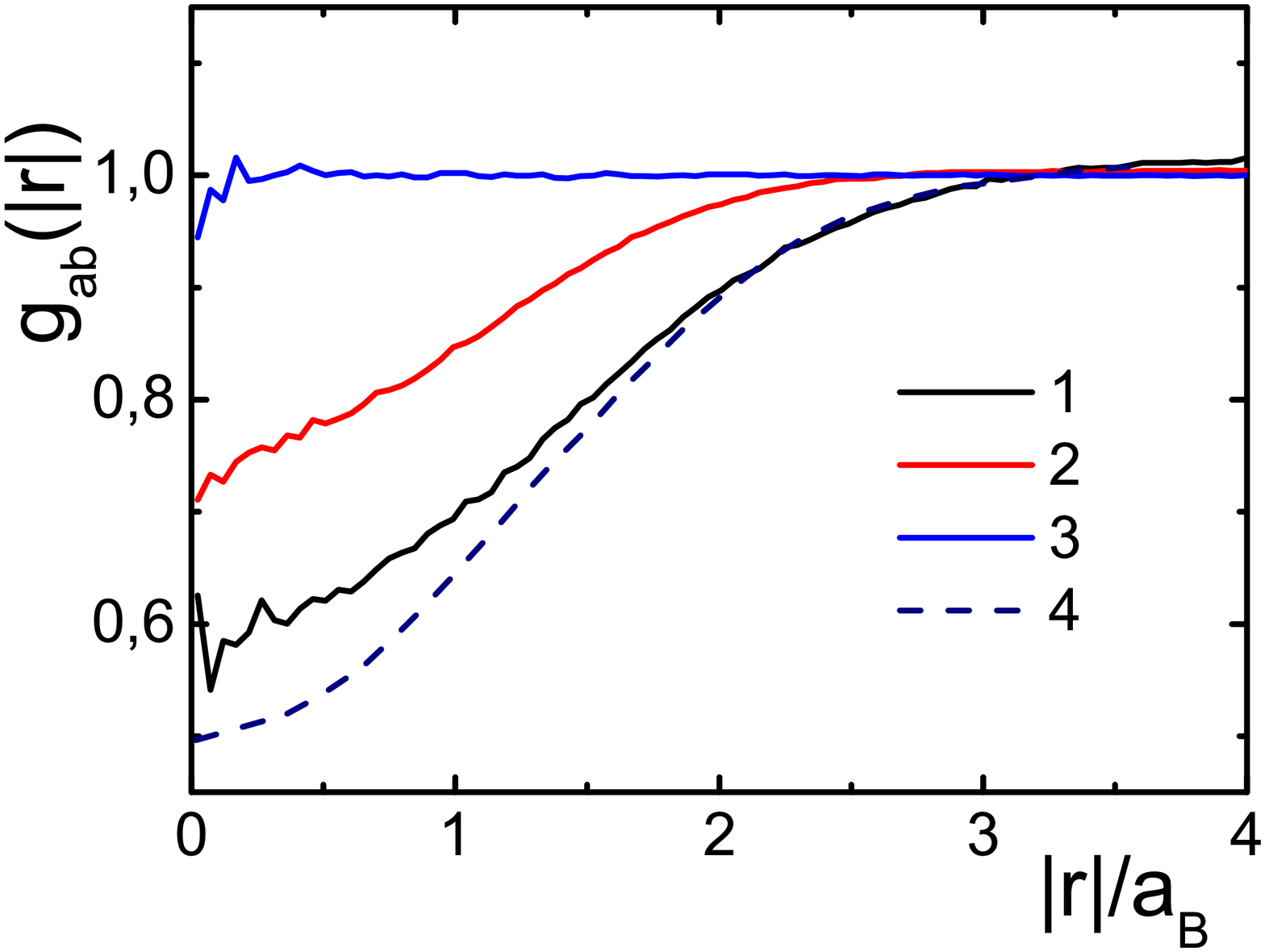}             
    \includegraphics[width=8.9cm,clip=true]{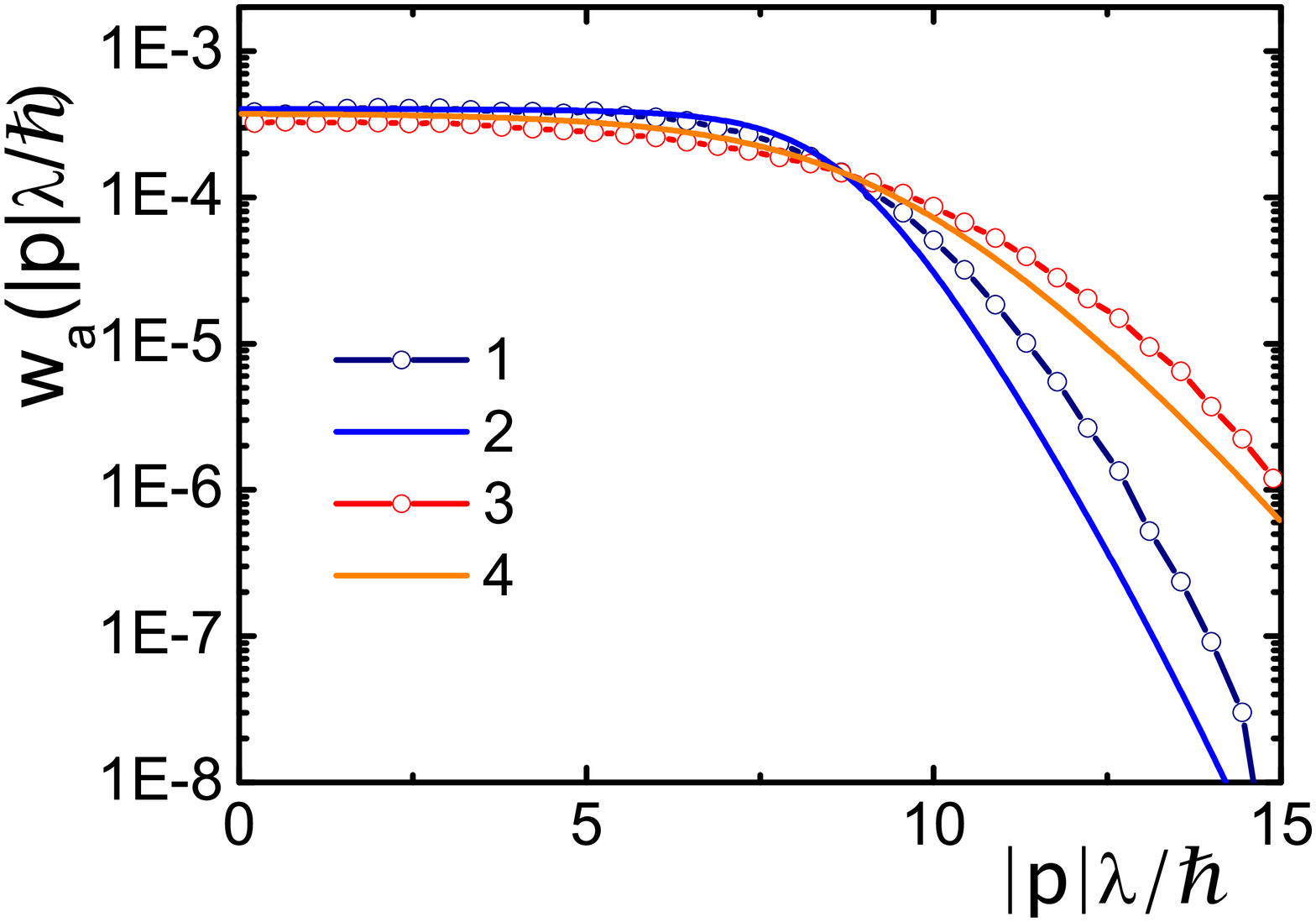}             
    \includegraphics[width=8.9cm,clip=true]{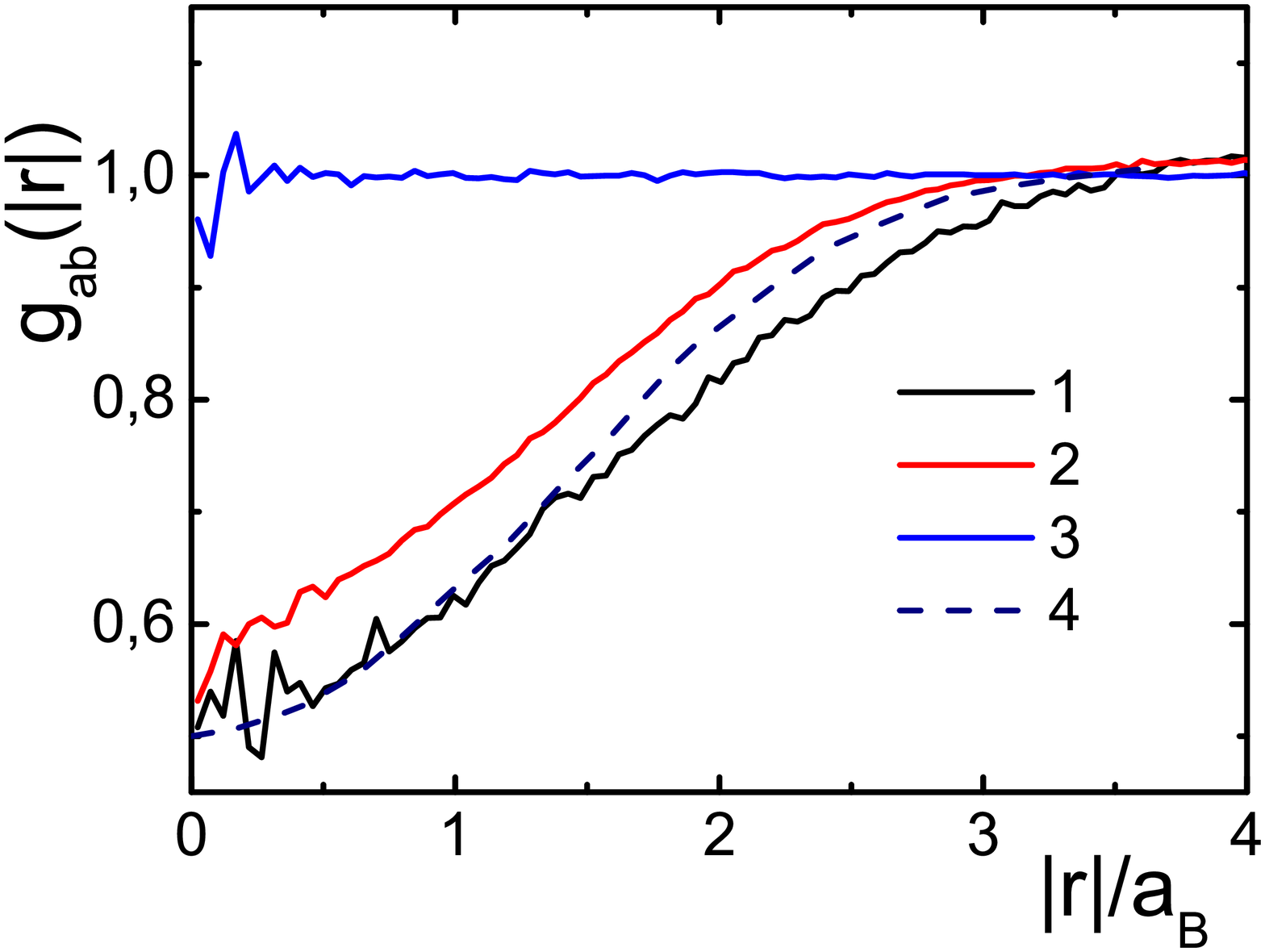}            
    \caption{(Color online) The momentum distribution functions $w_a(|p|)$ (left panels) and pair correlation functions 
        $g_{ab}(|r|),(a,b=e,h)$ (right panels) for ideal electron - hole plasma at $r_s=2$. 
        Left panels:
        lines $1,3$ show PIMC distributions $w_a(|p|)$  scaled by ratio of Plank constant to the electron thermal wavelength
        ($\frac{\hbar}{\lambda_e}$), while lines $2,4$ demonstrate the ideal Fermi distributions for electrons and two times heavier  holes respectively;
        Right panels:
        lines $1,2,3$ present PIMC electron -- electron, hole -- hole and electron -- hole  correlation functions scaled by Bohr radius respectively, 
        while line 4 show results of analytical approximations \cite{Bosse} for electrons. 
    }
    \label{DistrFunc1}
\end{figure}

Let us start from consideration of ideal fermions. 
For ideal electron -- hole plasma Figure~\ref{DistrFunc1} shows the
momentum distributions $w_{a}(|p_a|),(a=e,h)$ and
pair correlation functions $g_{ab}(|r|),(a,b=e,h)$ for
electron parameters of degeneracy equal to $n\lambda_e^3=5.6 \, (T/E_F=0.208, \, k_F \lambda_e=5.5) $ and 
$n\lambda_e^3=10 \, (T/E_F=0.141, \, k_F \lambda_e=6.66)$ for upper to bottom rows respectively. 
Let us note that holes is this calculations are two times  heavier than electrons, so the related parameters of degeneracy
is $2^{3/2}$ times smaller. Presented momentum distribution functions are normalized to one.
As it follows from the analysis  of figure~\ref{DistrFunc1}
agreement between PIMC calculations and 
analytical Fermi distributions (FD) 
is good enough up to parameter of electron degeneracy equal to 
$n\lambda_e^3=10 \, (T/E_F=0.141)$. Here integral characteristics such as energy
resulting  from PIMC and analytical distribution functions are
practically equal to each other.

In the right panels we compare with good agreement the calculated here pair correlation functions with 
analogous analytical approximations obtained in \cite{Bosse} for electrons.  At the distances 
lesser than thermal wave length of electron or hole
the influence of Fermi repulsion is evident with increasing
parameter of degeneracy from upper to bottom rows. At the same time
the electron - hole pair correlation functions are identically equal
to unit because exchange interaction between particle of different type is missing.

\begin{figure}[htb]
    \includegraphics[width=7.25cm,clip=true]{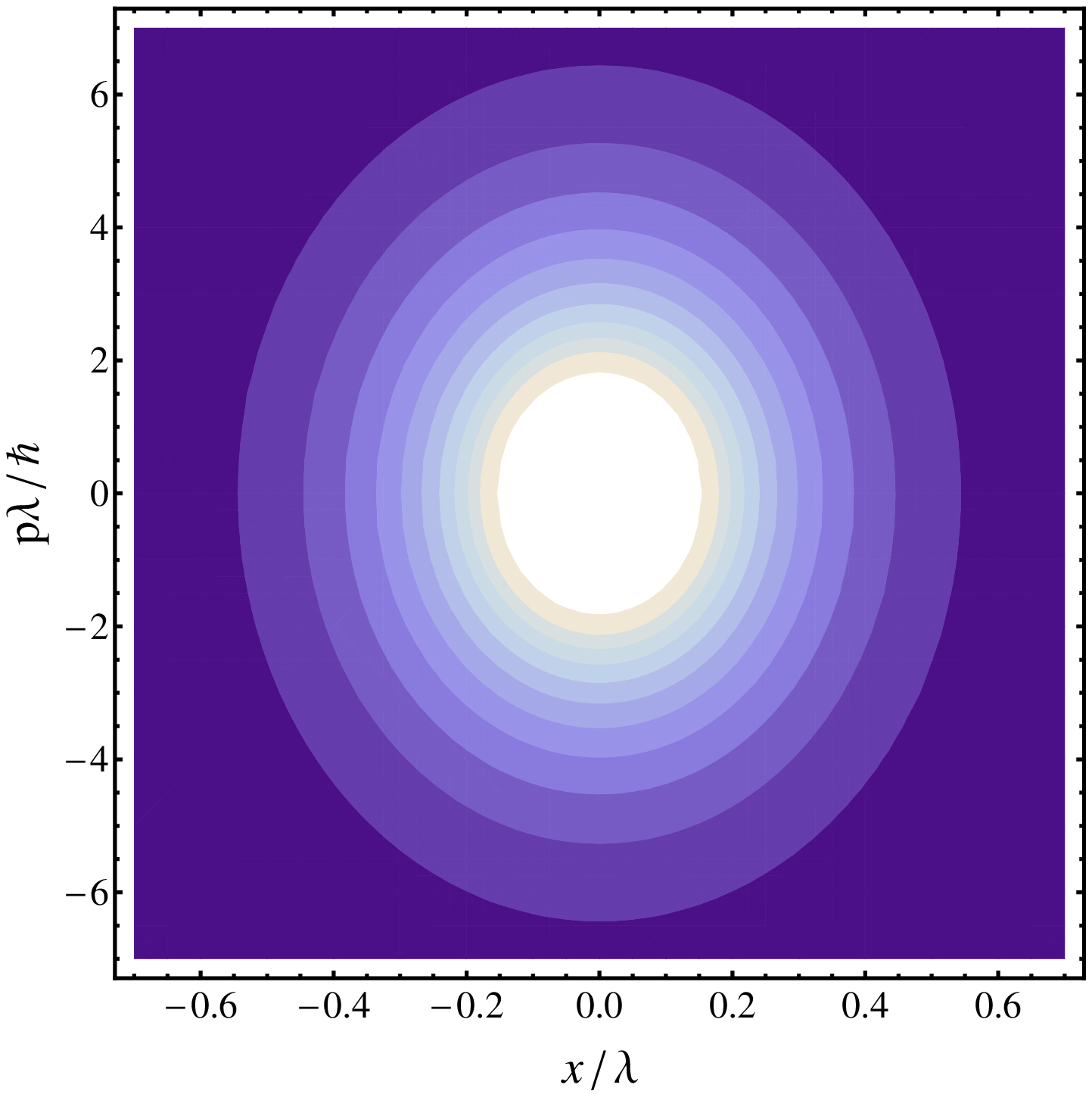}
    \hspace{1.3cm}
    \includegraphics[width=7.25cm,clip=true]{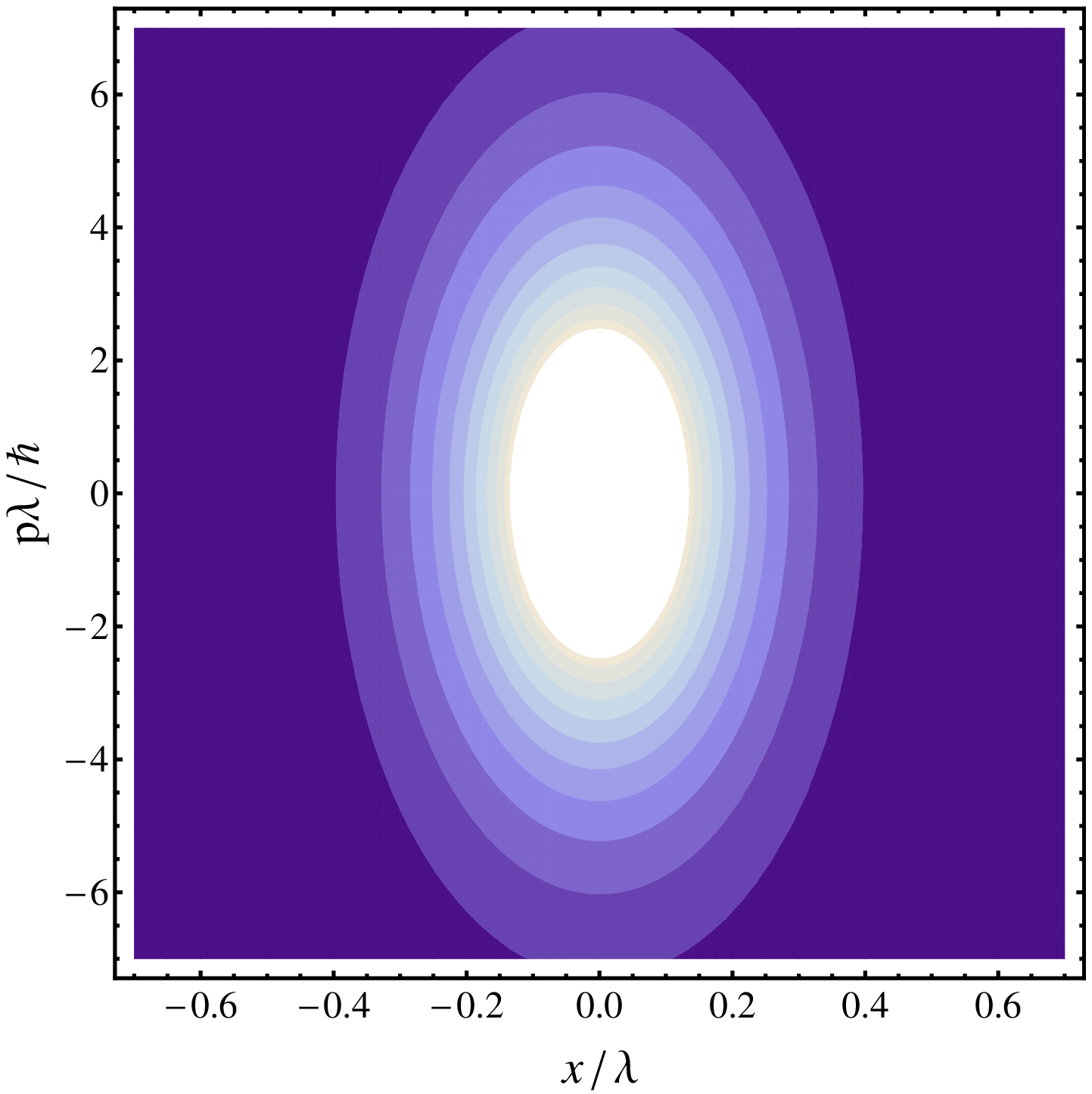}
    \caption{(Color online) Contour plots of the repulsive effective exchange pair  pseudopotentials in phase space.
        Left panel: $m_e=1$, dark area $\beta v^{e}_{ lt}\approx 0$, white area $\beta v^{e}_{ lt}\geq 1.9$.
        Right panel: $m_h=2$, dark area $\beta v^{h}_{ lt}\approx 0$, white area $\beta v^{h}_{ lt} \geq 1.5$.
    }
    \label{ExPt}
\end{figure}

Presented results for momentum distribution functions have been obtained in pair exchange approximation
described by effective pair pseudopotentials (\ref{vex}). Figure~\ref{ExPt}   presents contour
panels of exchange pair  pseudopotentials for parameter of degeneracy
equal to $5.6 \, (T/E_F=0.208)$. 
Momenta and coordinates  axises are scaled by the electron thermal wave length with Plank constant and factor ten for momentum.
As before holes are two times heavier than electrons.
As it follows from analysis  of Figure~\ref{DistrFunc1} the Pauli blocking in phase space accounted for by these exchange pseudopotentials
provides agreement of PIMC calculations and analytical FD in wide 
 ranges of fermion degeneracy and fermion momenta, 
where decay of the distribution functions is at least of five orders
of magnitude.
It necessary to stress that one of the reason of increasing discrepancy with degeneracy growth is limitation on
    available computing power allowing calculations with several hundred
    particles  each presented by twenty beads. When parameter of degeneracy is more than 
    $10 \, (T/E_F=0.141)$ the thermal 
    electron wave length is of order of Monte Carlo cell size and influence of finite number of particles
    and periodic boundary conditions becomes significant as has been tested in our calculations.

Let us stress that detailed test calculation for interacting particle carried out within developed approach in 
\cite{LarkinFilinovCPP,JAMP}  have demonstrated good agreement with results of independent calculations 
by standard path integral Monte Carlo method and available analytical results.  

\textbf{Quantum 'tails' in the momentum distribution functions}.
In classical statistics the commutativity of kinetic and potential
energy operators leads to the Maxwell distribution (MD) functions 
even for the strong interparticle interaction.
On the contrary for quantum systems the interparticle interaction
can affect the maxwellian shape of the particle kinetic energy
distribution function, because the
interaction of a particle with its surroundings restricts the volume
of configuration space and  results in an increase in the volume of the momentum space
due to the uncertainty principle, i.e., in a rise in the fraction of particles with higher
momenta. On other
hand Fermi statistical effects accounted by Pauli blocking of
fermions can be also important at the same conditions. On account of
these effects the momentum distribution functions may contain a
power-law tails even under conditions of thermodynamic equilibrium
\cite{Galitskii,Kimball,StarPhys,Eletskii,Emelianov,Kochetov,StMax}. 
\begin{figure}
    \includegraphics[width=8.9cm,clip=true]{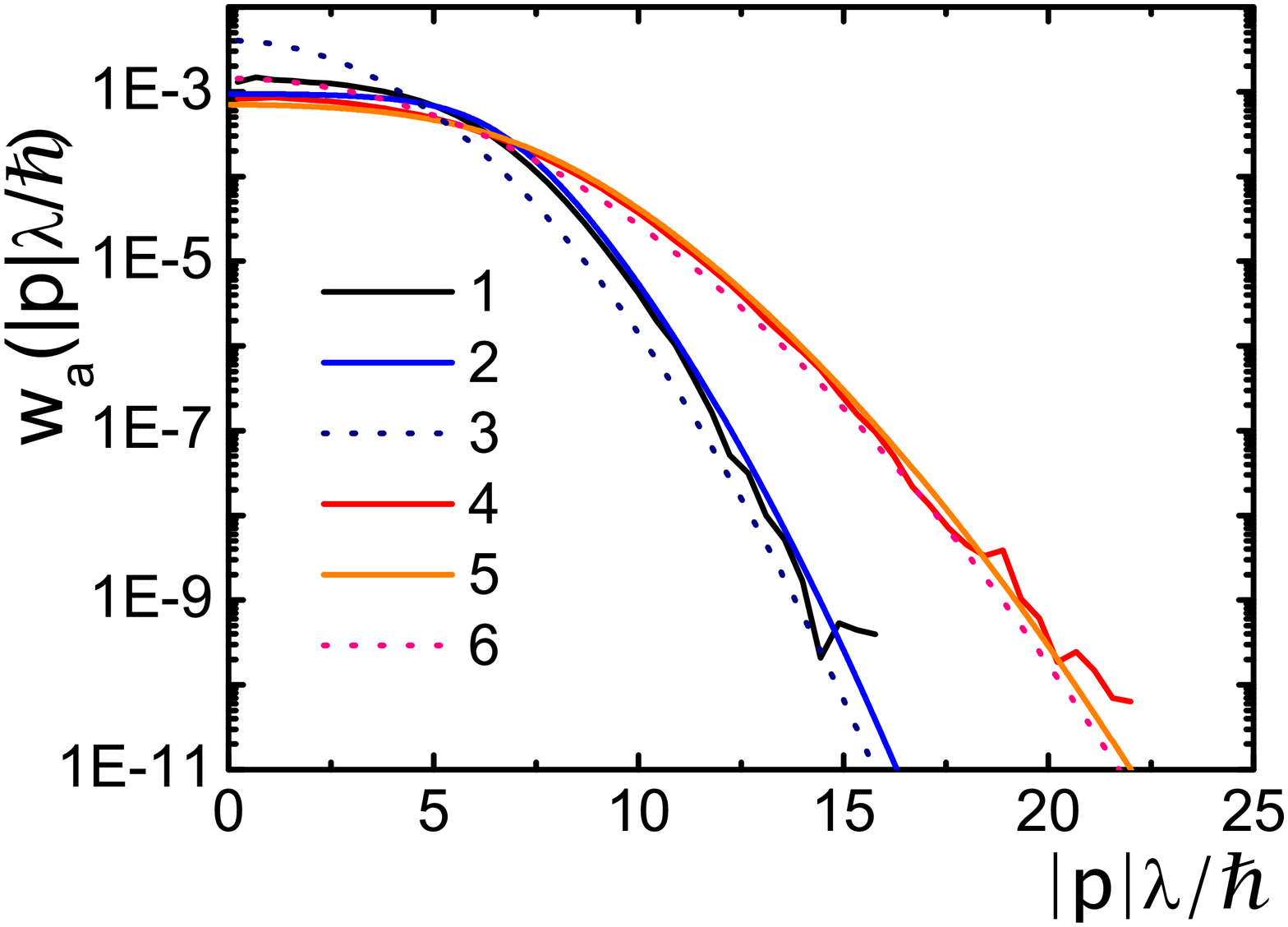}                       
    \includegraphics[width=8.9cm,clip=true]{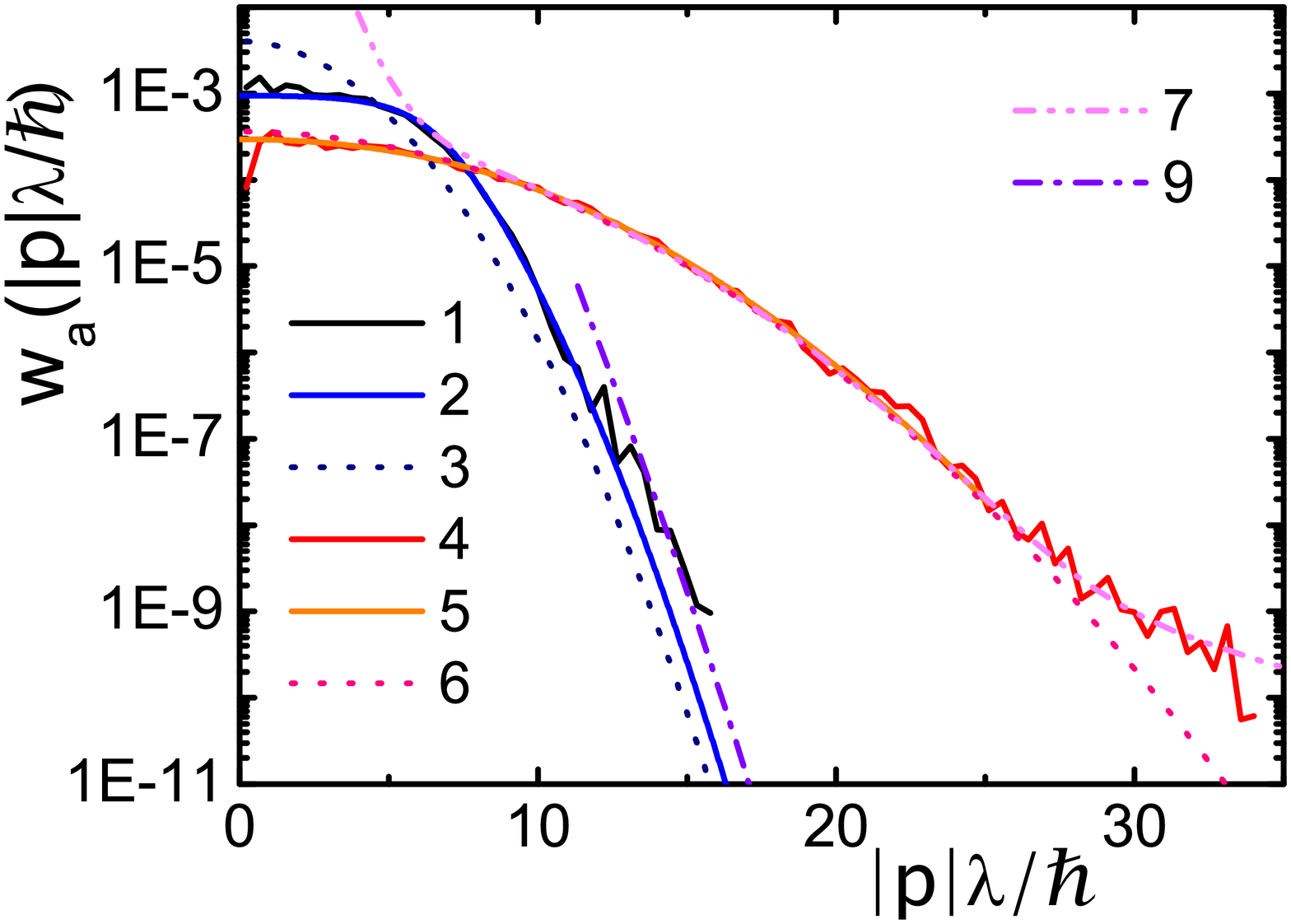}                    
    \includegraphics[width=8.9cm,clip=true]{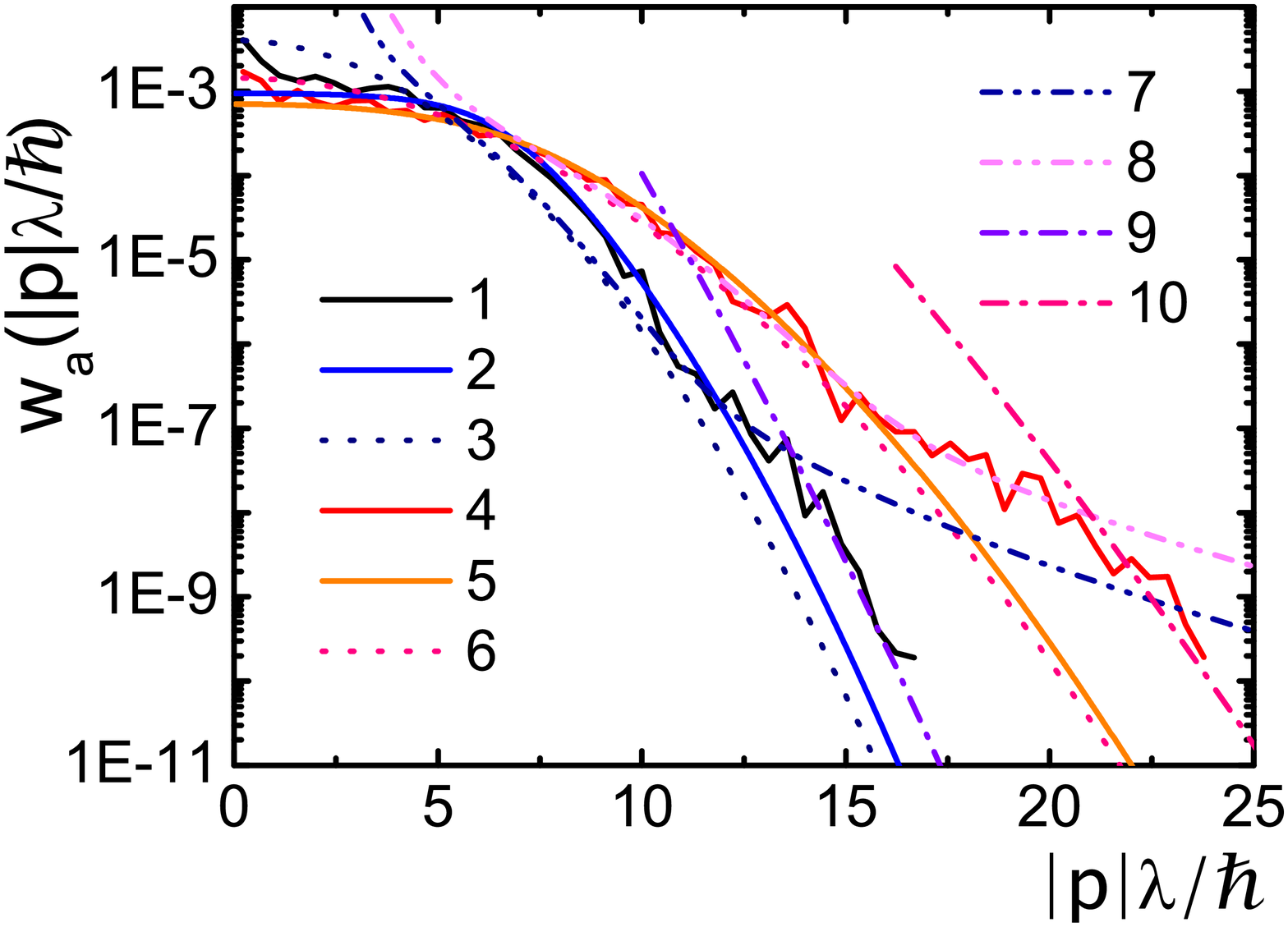}                       
    \includegraphics[width=8.9cm,clip=true]{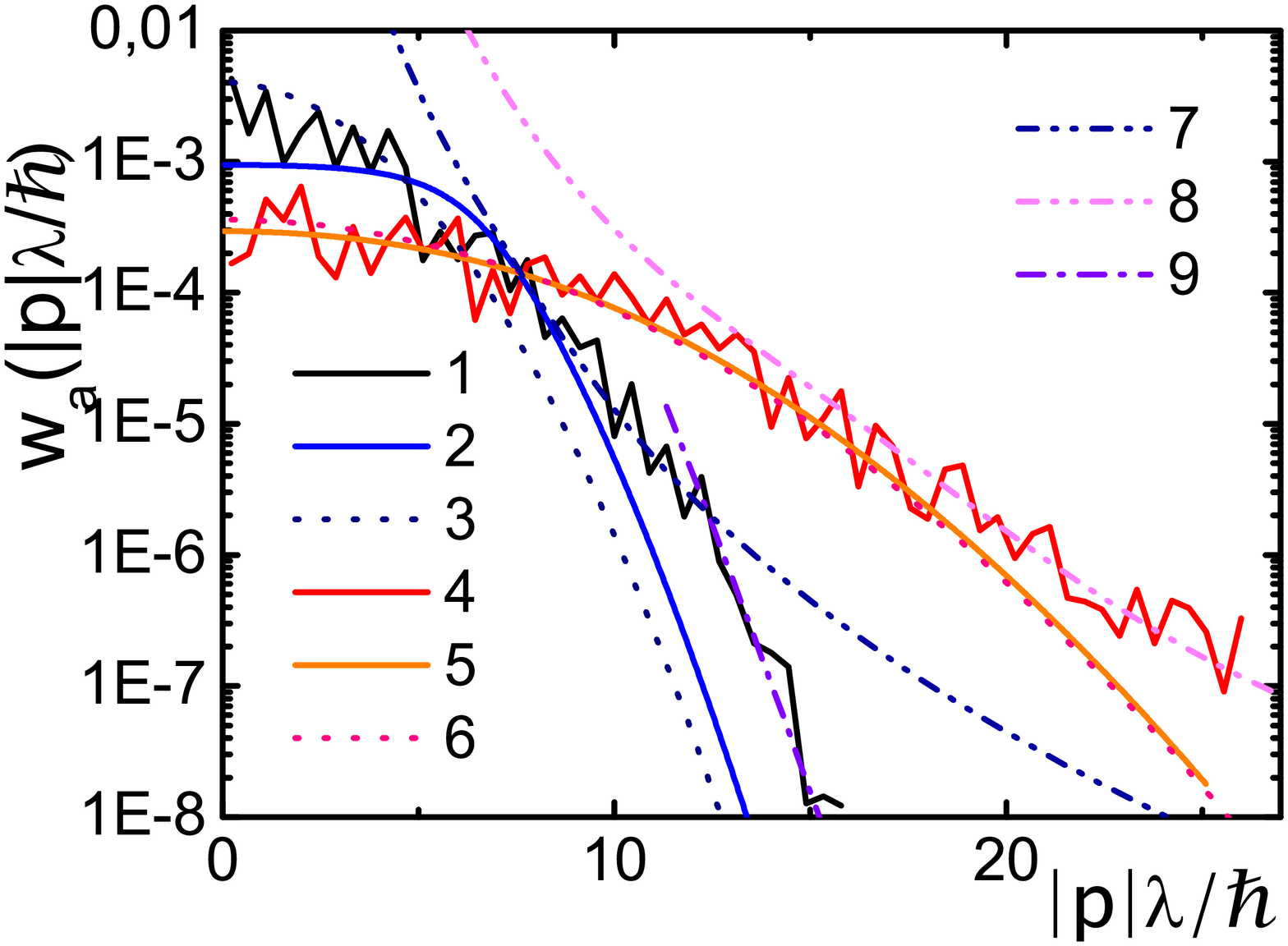}                     
    \includegraphics[width=8.9cm,clip=true]{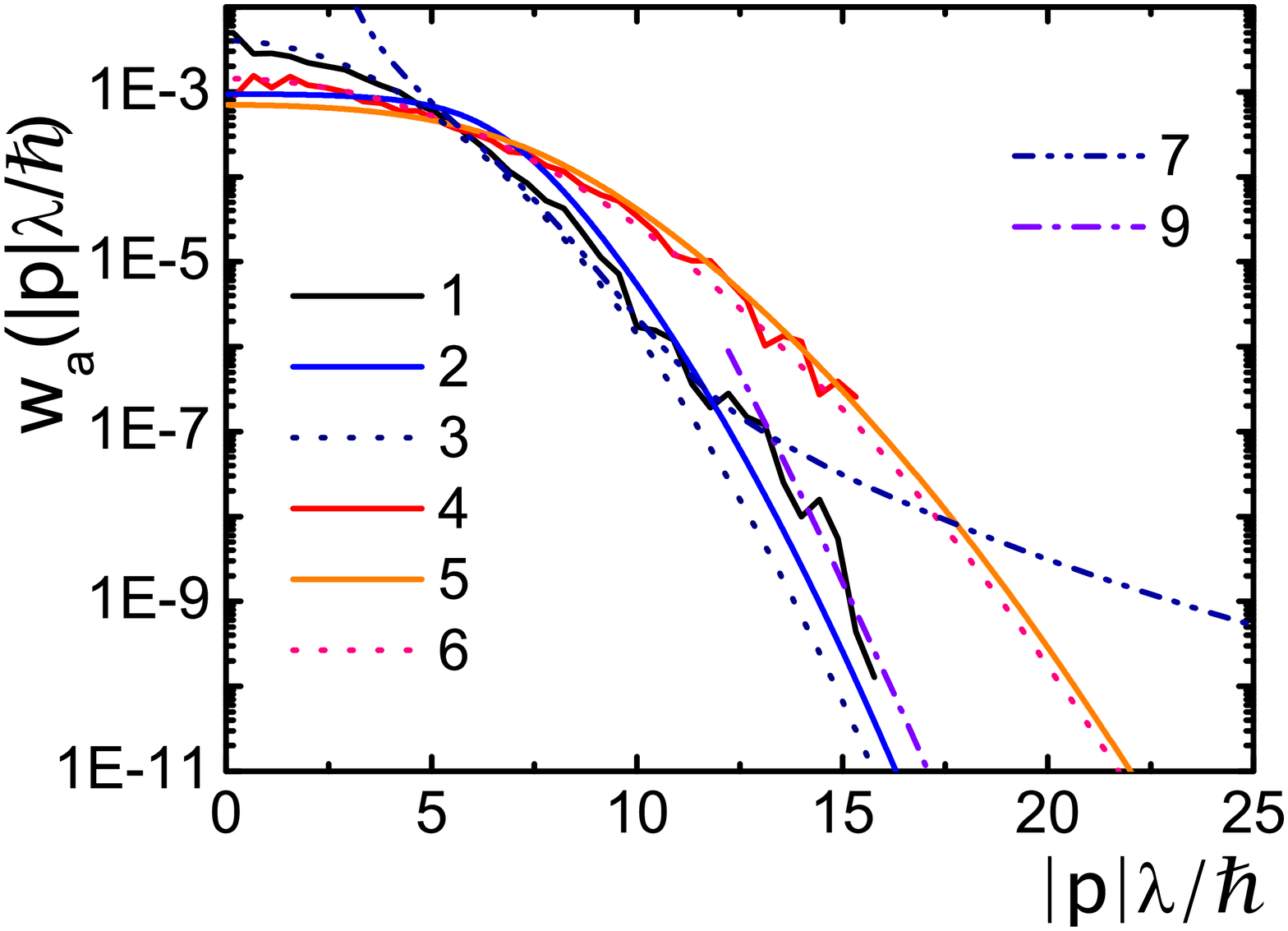}                       
    \includegraphics[width=8.9cm,clip=true]{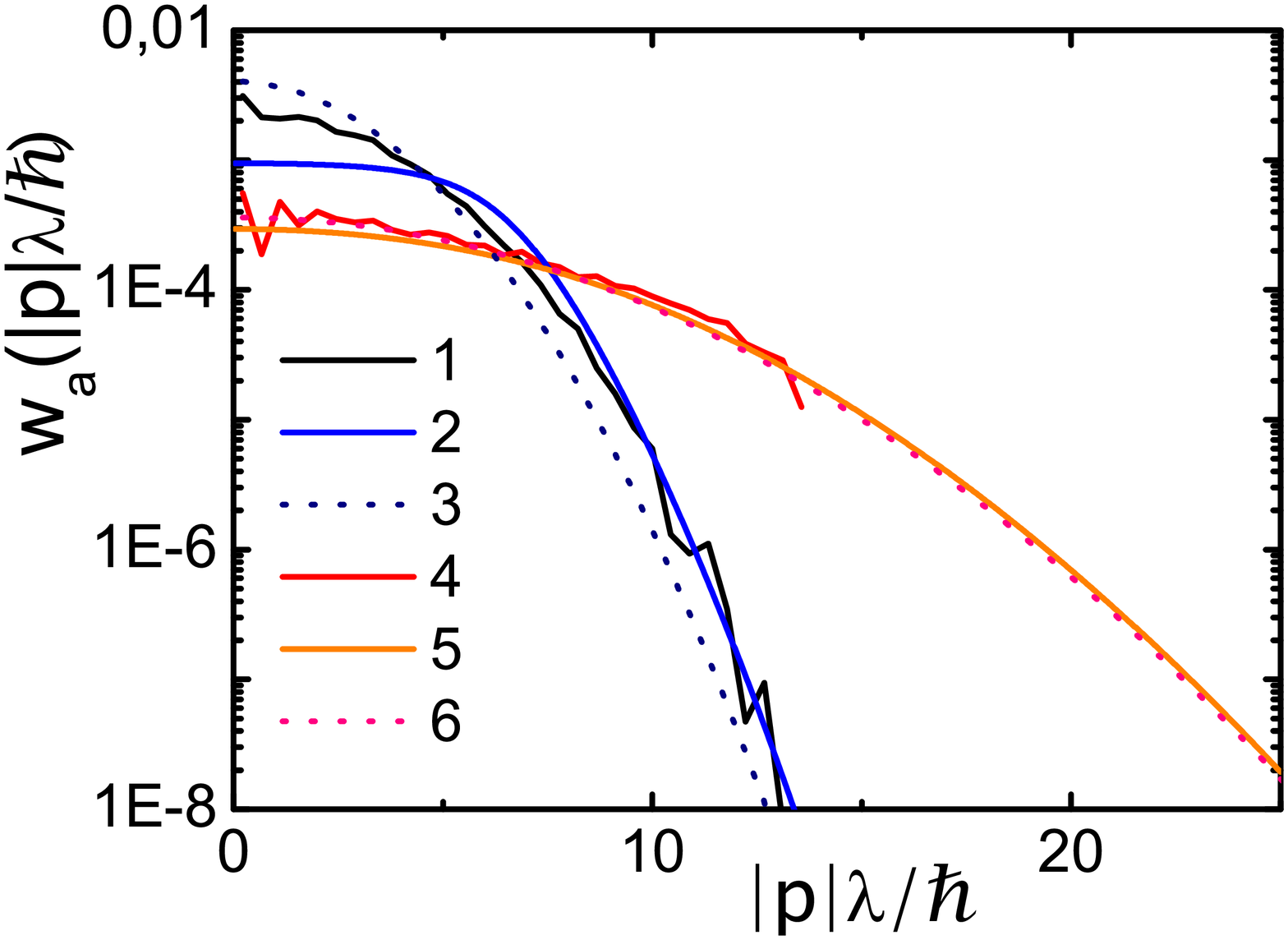}                     
    \caption{(Color online) 
        The momentum distribution functions $w_a(|p|)$  for non - ideal electron -- hole plasma.
        Results for electrons and two (five) times heavier holes are presented by left (right) panels respectively.
        Electronic momentum distributions are presented by lines (1,2,3,7,9), while analogous results for holes are presented by lines (4,5,6,8,10).
        Detailed description of the meaning of these lines is given in text.
        Degeneracy of electrons is equal to 4 ($T/E_F=0.261, \, k_F \lambda_e=4.91$), while $r_s$ and  plasma classical coupling parameter is equal 1, 2, 4 
        from upper to bottom rows respectively. Oscillations of PIMC distribution functions designate statistical errors.
    }
    \label{DistrFunc25}
\end{figure}
In these papers in frameworks of some models and perturbation theories it  was shown that for some system of 
non-relativistic particles the momentum distribution in the asymptotic region of large momenta 
$p$ may be described by the sum of MD and quantum correction proportional to $const/p^{8}$ (we use short notation as P8)
or MD with added product of $const/p^{8}$ and Maxwell distributions with effective
temperature that exceeds the temperature of medium (short notation -- P8MDEF) \cite{StMax}. Reliable calculation of 
numerical parameters in these asymptotic is problematic. 

PIMC calculations for the electron and hole momentum distribution
functions in electron -- hole plasma are presented  on Figure~\ref{DistrFunc25} by lines 1 and
4 for electrons and holes respectively. Here parameter of electronic degeneracy is equal to 
four  ($T/E_F=0.261$) and hole masses in two (left panels) and five
(right panels) times larger than electron mass. Figure~\ref{DistrFunc25} presents also several 
related analytical dependencies: FD (lines 2 and 5), MD (lines 3 and 6), P8 (lines 7 and 8),
P8MDEF (lines 9 and 10).
\begin{figure}[htb]
    \includegraphics[width=8.9cm,clip=true]{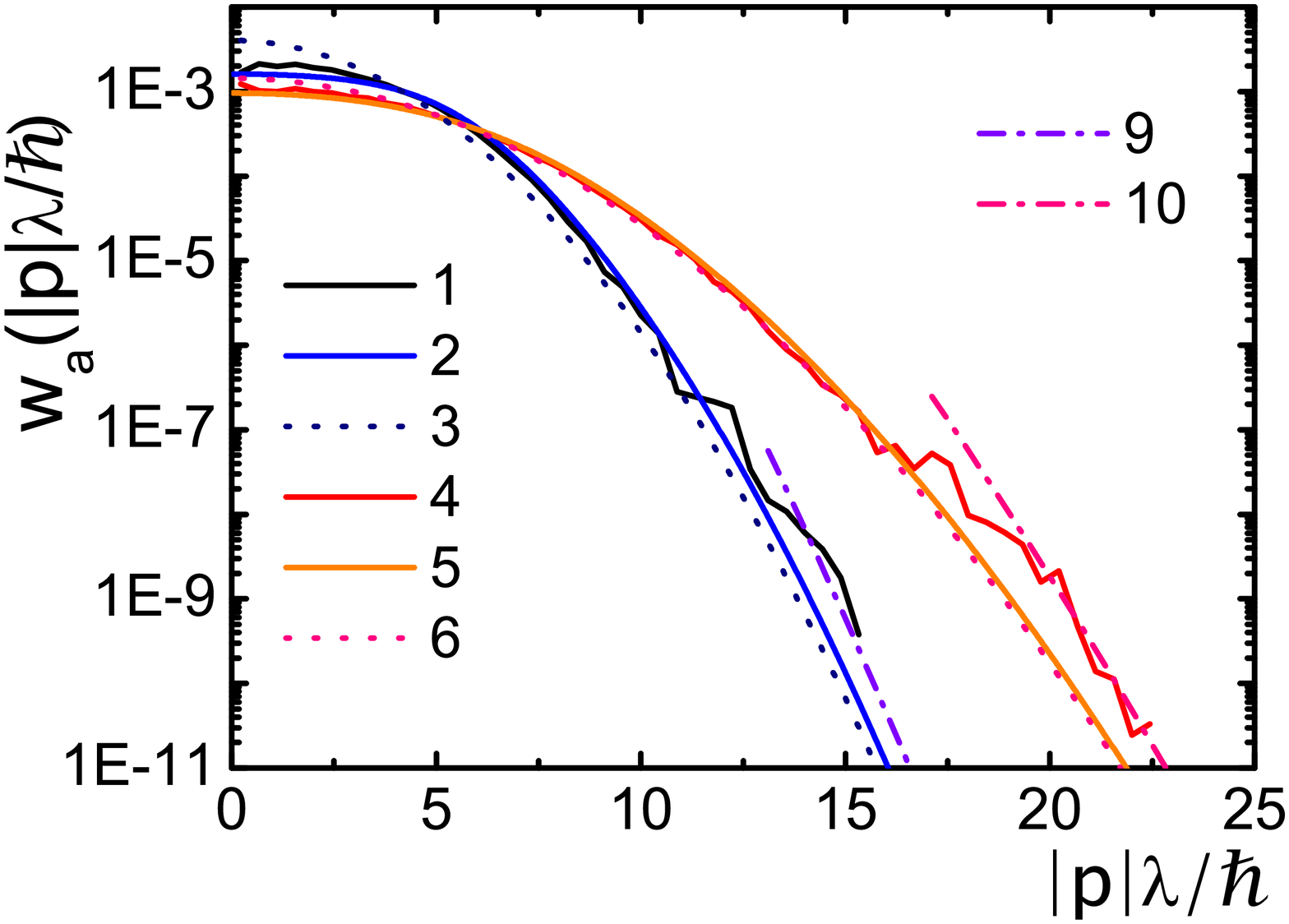}                     
    \includegraphics[width=8.9cm,clip=true]{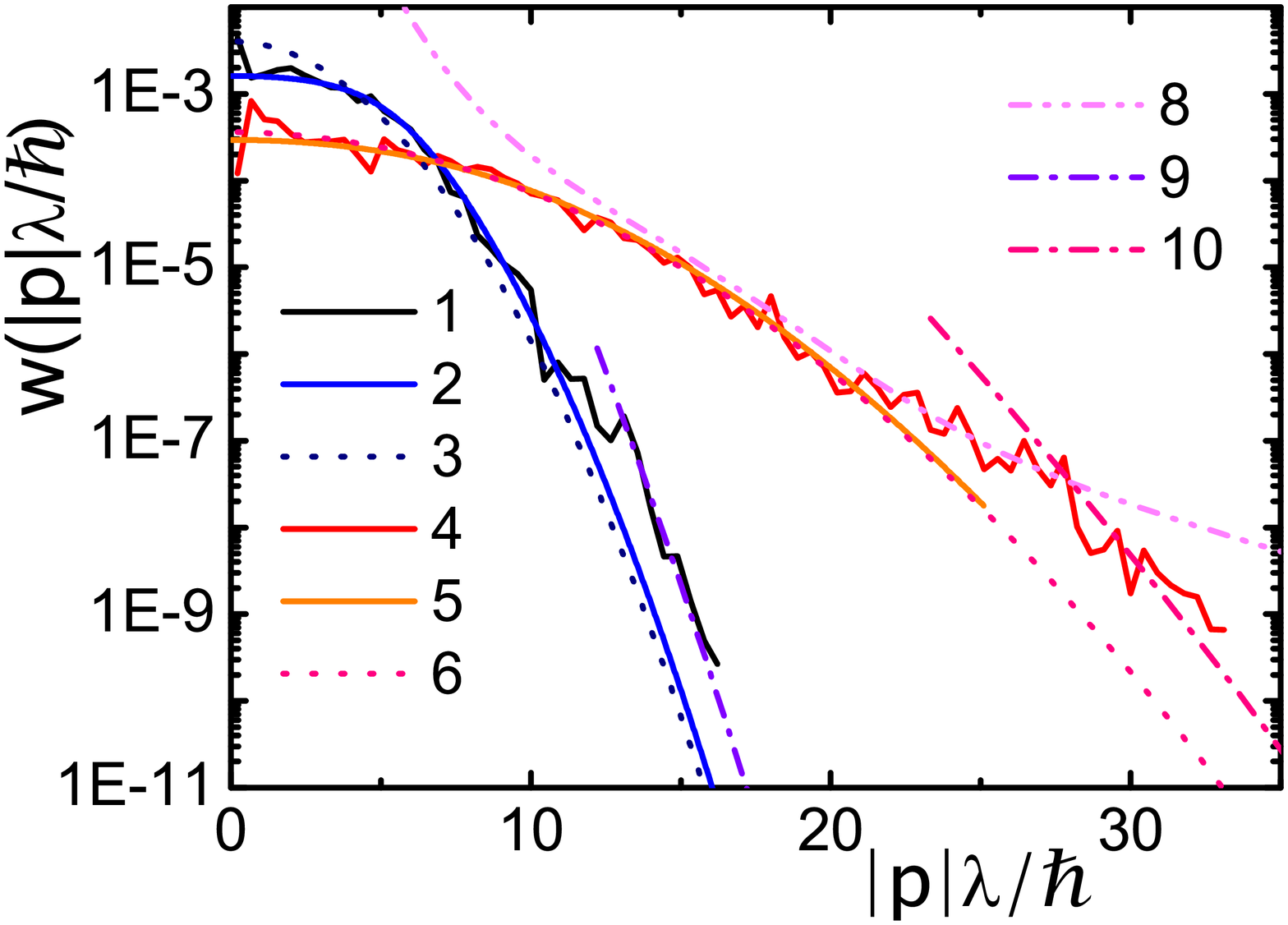}                     
    \includegraphics[width=8.9cm,clip=true]{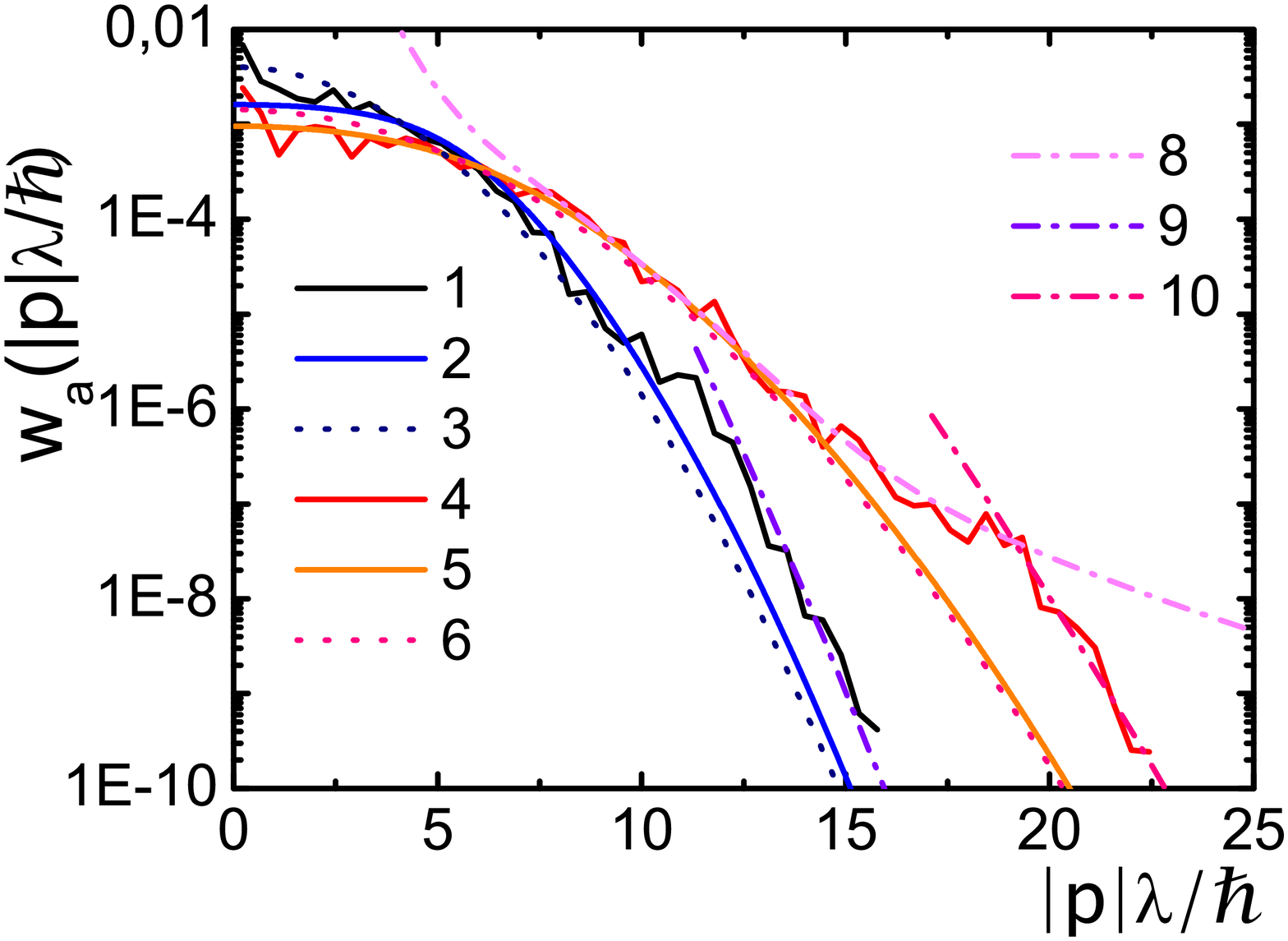}                         
    \includegraphics[width=8.9cm,clip=true]{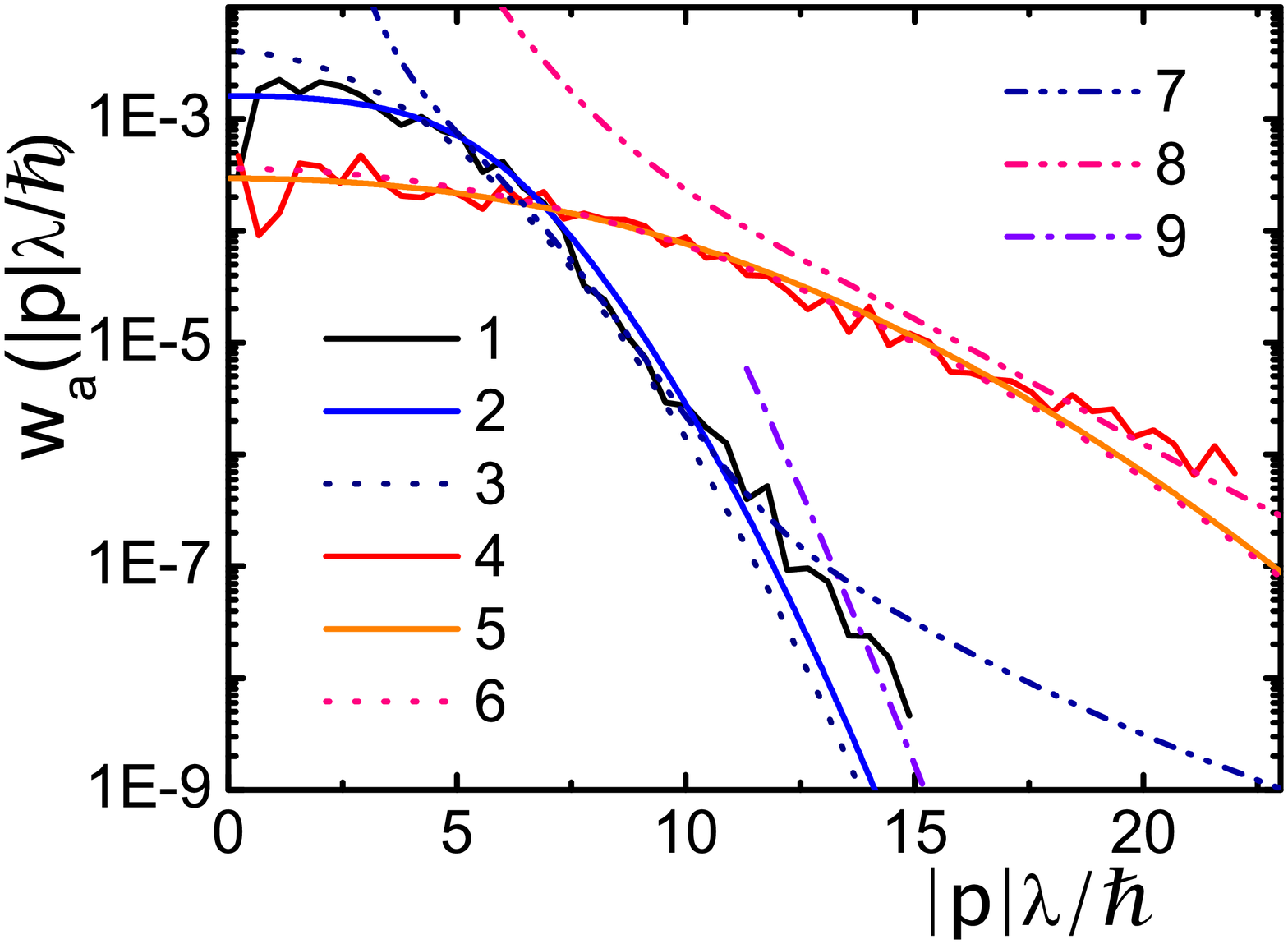}                        
    \includegraphics[width=8.9cm,clip=true]{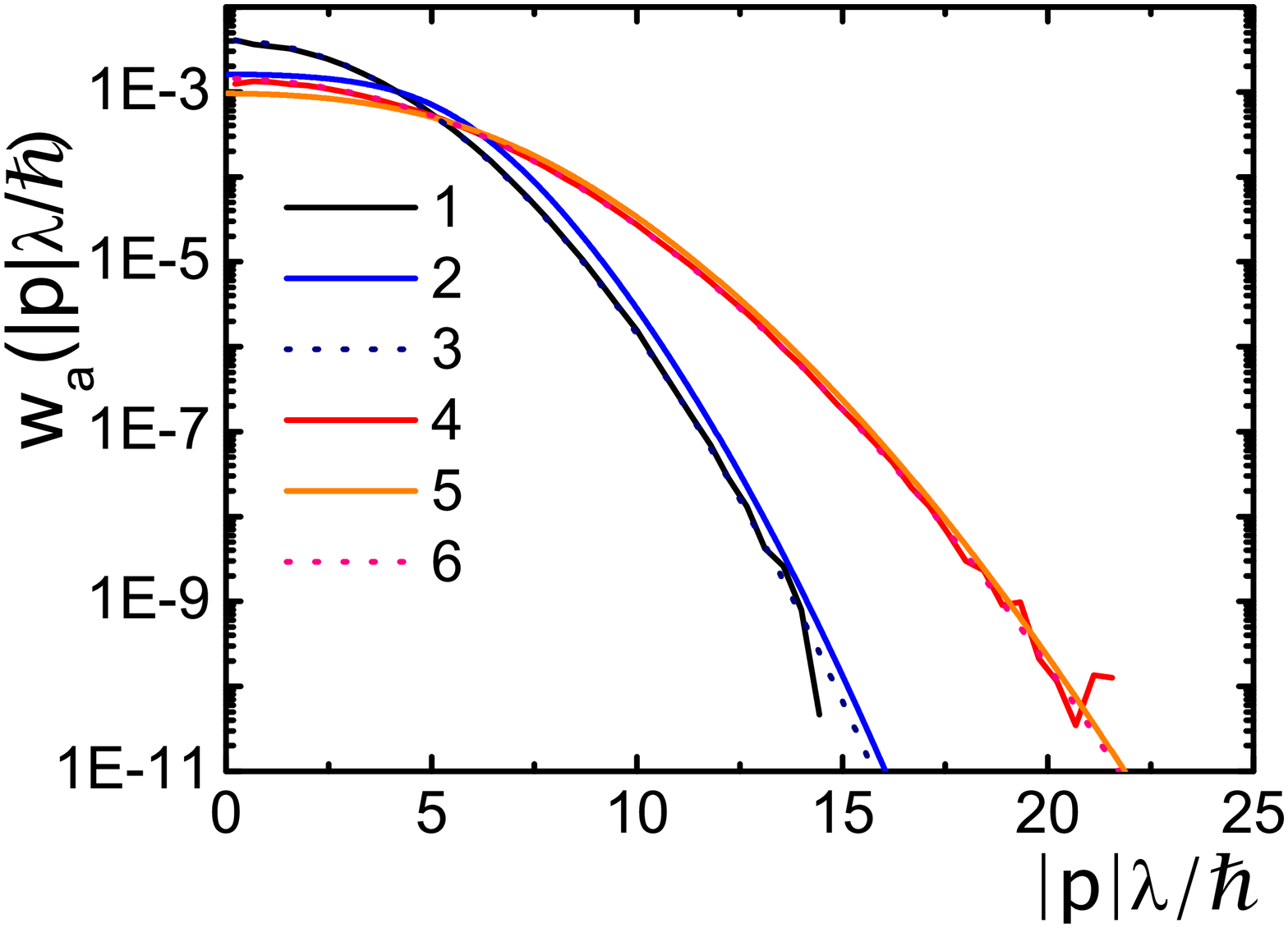}                   
    \includegraphics[width=8.9cm,clip=true]{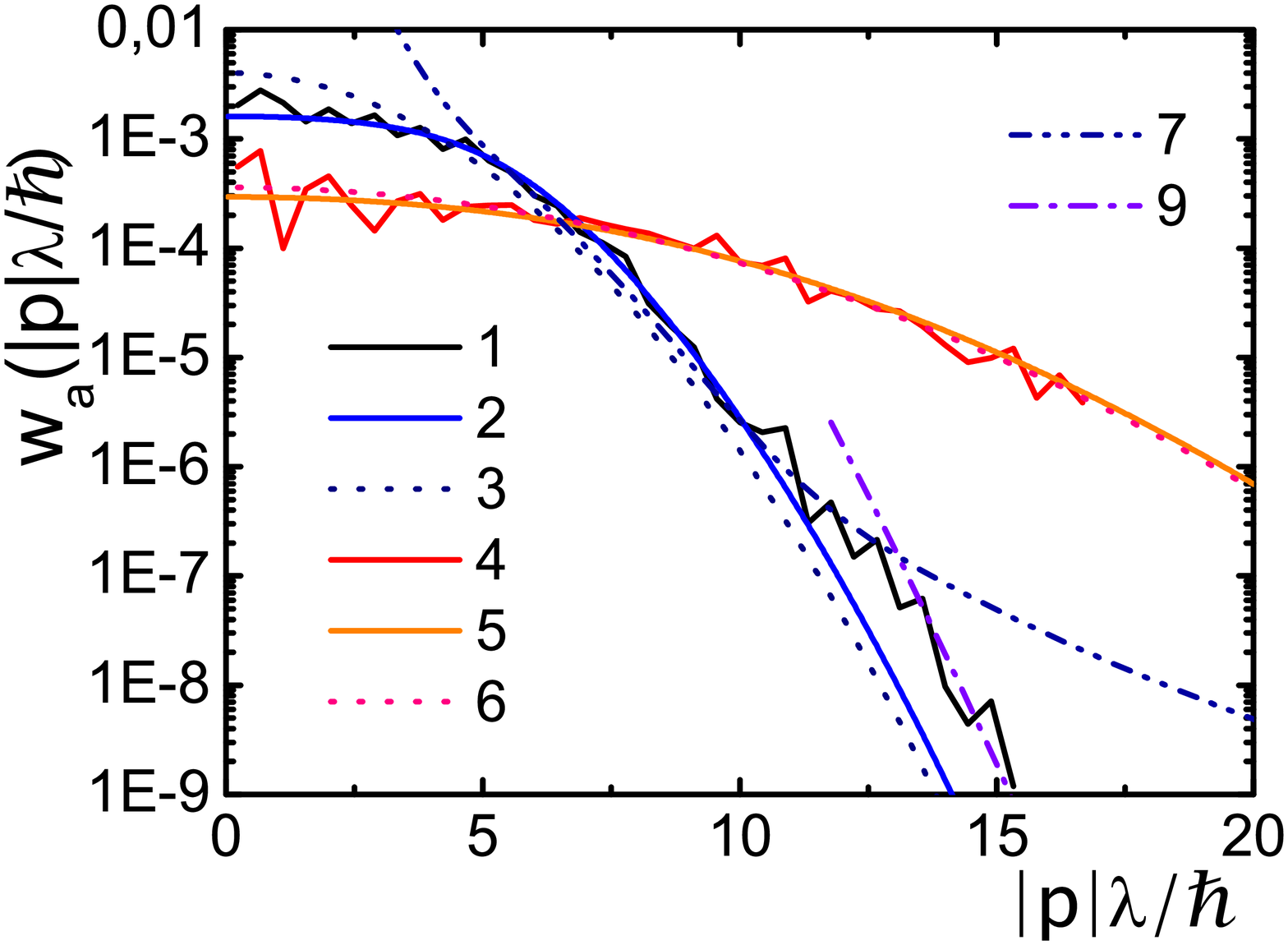}                     
    \caption{(Color online)
        The momentum distribution functions $w_a(|p|)$  for non - ideal electron -- hole plasma.
        Results for electrons and two (five) times heavier holes are presented by left (right) panels respectively.
        Electronic momentum distributions are presented by lines (1,2,3,7), while results for holes are presented
        by lines (4,5,6,8,9).
        Detailed description of the meaning of these lines is given in text.
        Degeneracy of electrons is equal to 2 ($T/E_F=0.413, \, k_F \lambda_e=3.9$), while  $r_s=2, \, 3, \, 4 $ and plasma classical coupling parameter 
        is equal 1.3, 2, 2.6 from upper to bottom rows. 
        Oscillations of PIMC distribution functions designate statistical errors.
    }
    \label{DistrFunc2}
\end{figure}

Here constant in P8 and effective temperature in P8MDEF have been
used as adjustable parameters to fit PIMC momentum distribution
functions in large momentum asymptotic regions. All MD and FD are normalized to one.
As it follows from analysis of Figure~\ref{DistrFunc25} dependence P8 may reliably
fit PIMC distributions in intermidiate region between FD
and P8MDEF,  
while the P8MDEF fits PIMC at large  momenta on almost all 
panels of Figure~\ref{DistrFunc25}.

It is necessary to stress that each type of asymptotics fits PIMC
momentum distribution function throughout decaying region of two or
even more order of magnitude. Interesting is the non monotonic
behavior of discrepancy between PIMC momentum distributions and FD or MD in
asymptotic regions with increasing classical coupling parameter
$\Gamma = \beta e^2 / (3/4\pi n_a)^{1/3} $ from top ($\Gamma =1$) to
bottom ($\Gamma =4$) panels. Namely for electrons and holes at low
$\Gamma \approx 1$ and high enough $\Gamma \approx 4$ the PIMC
distributions are approaching the FD or MD better than at $\Gamma
\approx 2$ or $\Gamma \approx 3$.

Physical meaning of this phenomenon is the following. At small $\Gamma \le 1$ the influence of
interparticle interaction is negligible and PIMC momentum distribution functions due
to the exchange pseudopotentials ($v^{a}_{ lt}\approx 2$) approximate related Fermi distributions.
At large $\Gamma \ge 4$ average interparticle interaction is more important than weaker 
exchange interaction and PIMC momentum distributions are approaching MD, which are very
close to FD at large momenta. This means that appearance of quantum 'tails' in intermediate
region of $\Gamma \approx 2,3$ may be explained by simultaneous influence of Coulomb and
interparticle exchange interactions.

\begin{figure}[htb]
    \includegraphics[width=8.9cm,clip=true]{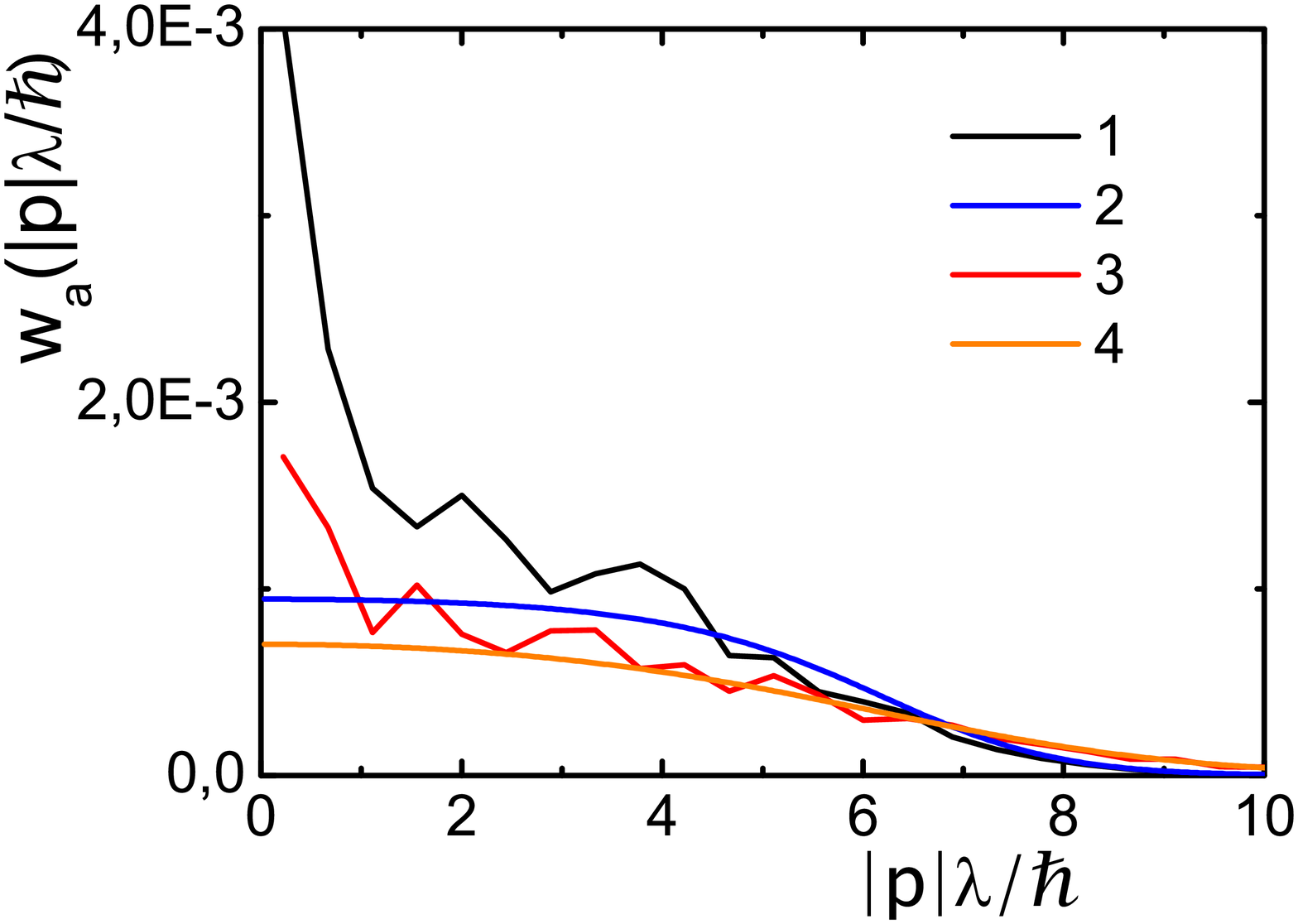}                         
    \includegraphics[width=8.9cm,clip=true]{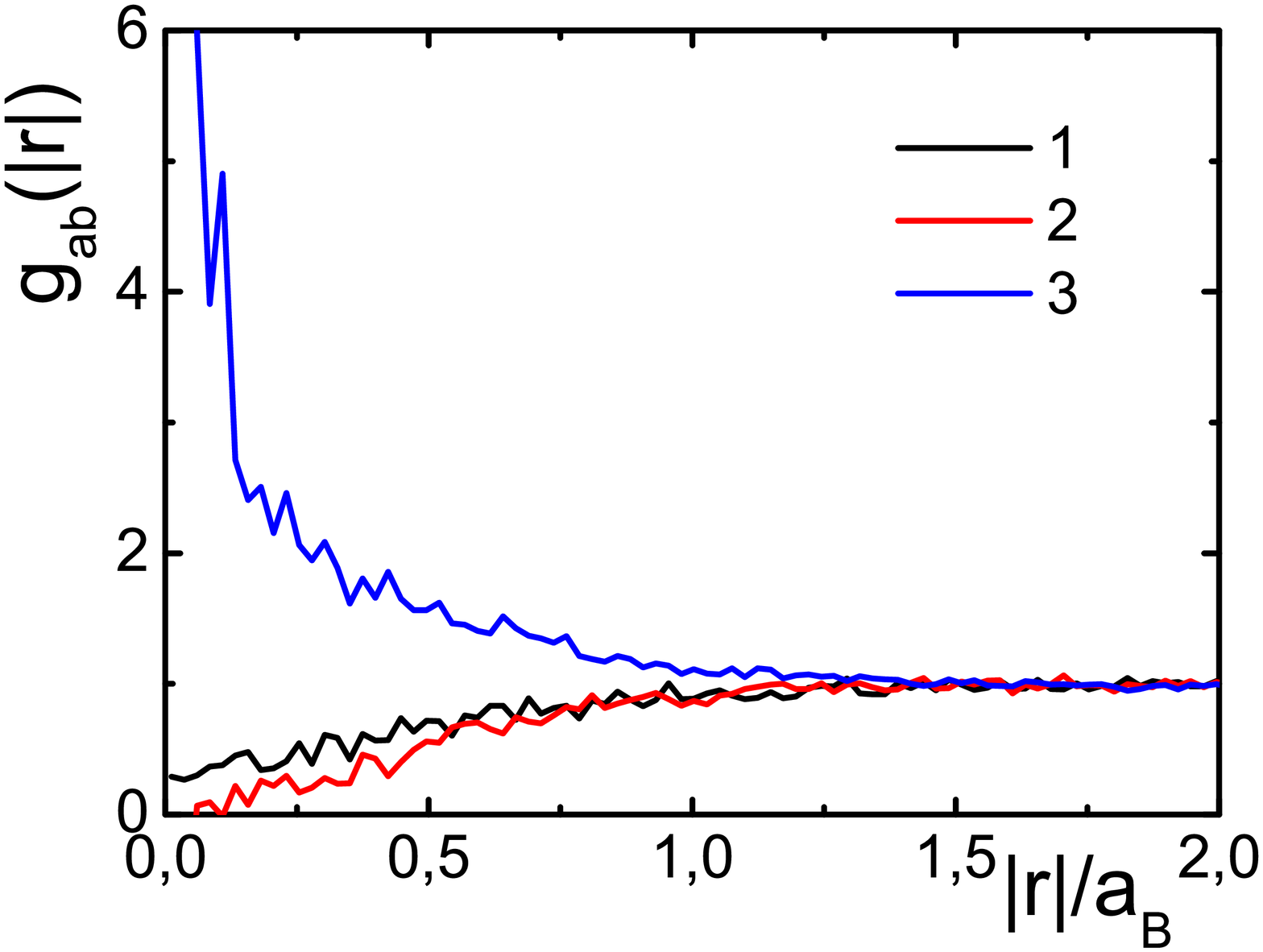}                      
    \includegraphics[width=8.9cm,clip=true]{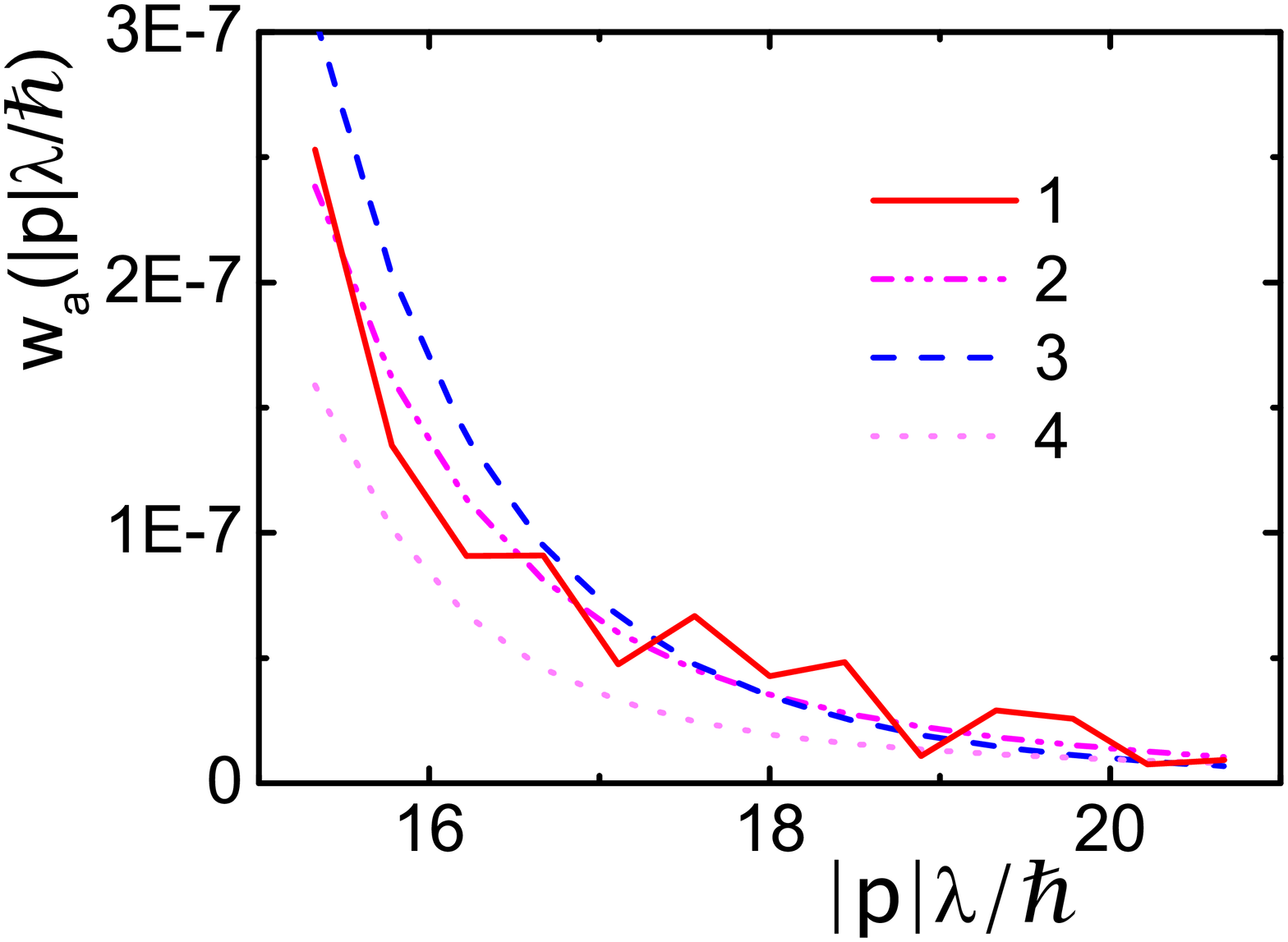}                        
    \includegraphics[width=8.9cm,clip=true]{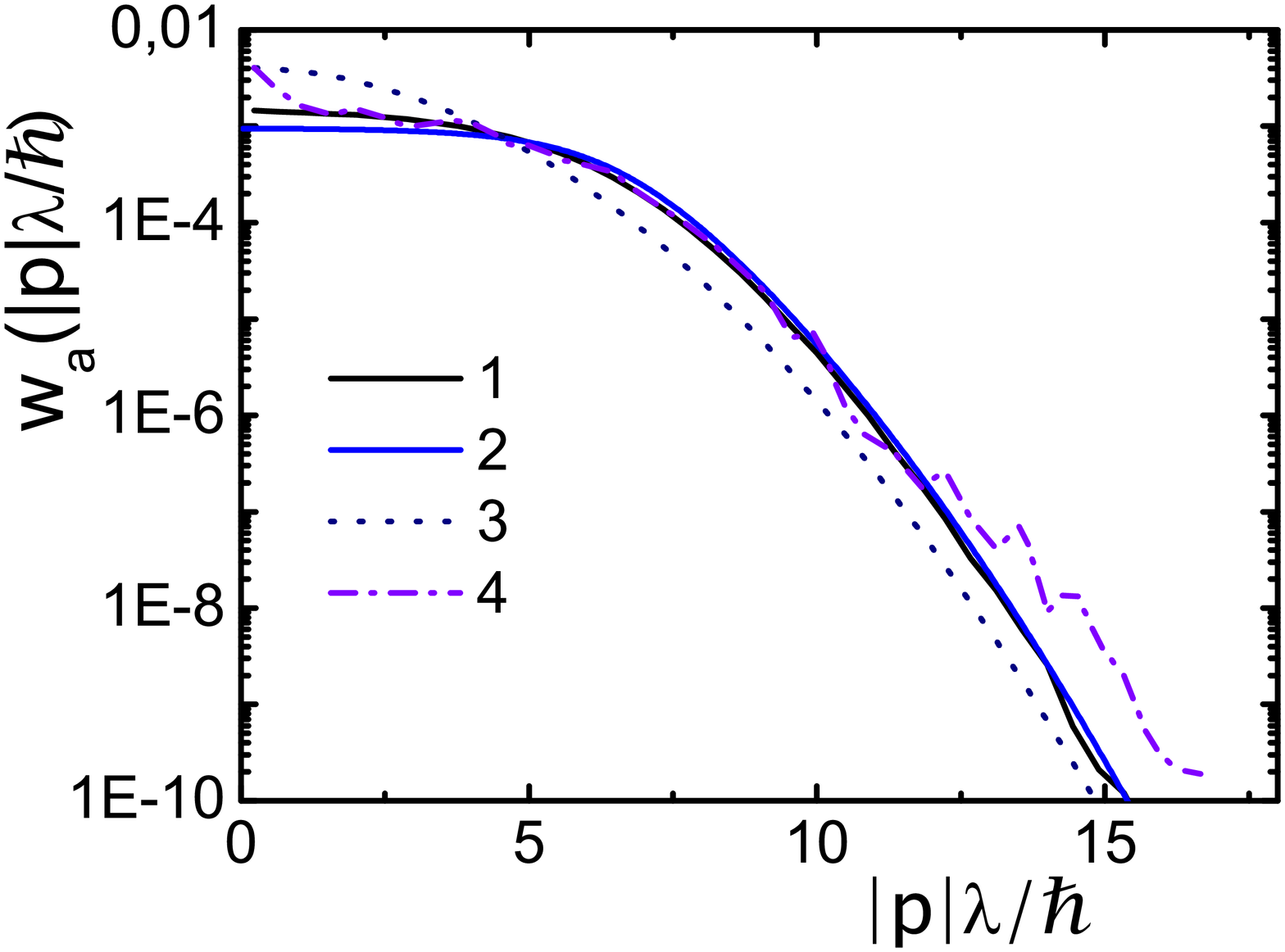}                        
    \caption{(Color online) Momentum distributions and pair correlation functions for electrons and two times heavier hole at 
    degeneracy of electrons equal to 4 ($T/E_F=0.261, \, k_F \lambda_e=4.91$). \\
        Top left panel: $w_a(|p|)$ at $p\lambda/\hbar\le 10$ and $r_s=2$.
        Lines: 1,2 -- PIMC and FD for electrons; 3,4 -- PIMC and FD for holes respectively. \\
        Top right panel: $g_{ab}(|r|),(a,b=e,h)$ at $r_s=1$ .  
        Lines 1,2,3 present electron - electron, hole - hole and electron - hole distribution  functions respectively.
        Plasma classical coupling parameter is equal $1$. 
        Small oscillations of PIMC distribution functions designate statistical error. \\
        Bottom left panel: power-low approximations $const/p^n$ (P8) of PIMC momentum distribution in intermediate region at $r_s=2$. 
        Lines: 1 -- PIMC momentum distribution; 2 -- $const/p^n$ for $n=8$; 3 -- $const/p^n$ for $n=11$; 4 -- $const/p^n$ for $n=8$. \\
        Bottom right panel: Lines: 1 -- PIMC for UEG; 2 -- FD; 3 -- MD; 4 -- PIMC for electron -- hole plasma. 
        Classical coupling parameter $\Gamma$ equal to two,  $r_s=2$   
    }
    \label{PowerLow8}
\end{figure}

Figure~\ref{DistrFunc2}  shows PIMC distribution functions for
parameter of electronic degeneracy equal to two ($T/E_F=0.413$) 
and $m_h=2$ (left panels) and  $m_h=5$ (right panels). Presented results
confirm even more clearly the basic features of asymptotic behavior
of PIMC distribution functions discussed above.

Figure~\ref{PowerLow8} presents PIMC momentum distributions and pair correlation functions.
At the top left panel  at momenta ($p\lambda/\hbar\le 10$)
the PIMC distributions show systematic oscillations and exceeding of FD
for both electrons and holes, which is consequence of 
interparticle interaction resulting in partial ordering of electrons
and holes as can be seen from right panel.
Typical particle configurations are described by the pair correlation functions presented
by the top right panel. 
This panel shows results of calculation of pair correlation
functions  for non ideal electron -- hole plasma at $\Gamma=1$. On the
contrary to ideal plasma the electron -- hole pair correlation
function increases at the small interparticle distance due to the
Coulomb attraction, while electron - electron and hole - hole pair
correlation function decreases due to the Coulomb and Fermi
repulsion. Coulomb electron - electron repulsion is
lesser pronounced in comparison with hole -- hole
repulsion due to tunneling effects, which are
larger for lighter electrons. Here the electron - electron  Fermi
repulsion is lesser pronounced in comparison with
ideal plasma, because the coupling parameter is strong enough
(compare this function with related $g_{ee}$ on Figure~\ref{DistrFunc1}).

Bottom left panel of Figure~\ref{PowerLow8} presents approximation of PIMC momentum
distribution function in intermediate region by several power -- low
dependences with different values of power.
Analysis of figure~\ref{PowerLow8} clearly shows validity of
approximation described by dependence P8.

To stress the role of interparticle interaction and correlation
effects we have carried out calculations of electron momentum
distribution functions for the model of uniform
electronic gas (UEG) being the quantum mechanical
model of interacting electrons where the positive charges are
assumed to be uniformly distributed in space. Momentum distribution
functions for UEG model and electron -- hole plasma are presented by
bottom right panel of Figure~\ref{PowerLow8} for parameter of degeneracy
equal to four ($T/E_F=0.261$), $r_s=2$ and classical coupling parameter equal to 2.
(the same as for Figure.~\ref{DistrFunc25}).  Comparison of obtained results
demonstrates disappearance of quantum 'tail' and very good agreement
of electron momentum distribution functions of UEG  with Fermi
distribution. Let us stress that studies of momentum distribution functions of UEG by developed  
approach are at present in progress, while studies of thermodynamic properties of UEG 
by PIMC and detailed comparison with available data (see for example \cite{UEG1}) have been recently 
published by the authors of this paper in \cite{UEG}.

\section{Conclusion}

In our work we have derived the new path integral representation of
Wigner function of the Coulomb system of particles for canonical ensemble.
We have obtained explicit expression of Wigner function in  linear 
approximation resembling the Maxwell --  Boltzmann distribution on momentum variables, but
with quantum corrections.
This approximation contains also the oscillatory multiplier describing quantum
interference.
We have developed new quantum Monte-Carlo method for calculations
of average values of arbitrary quantum operators depending both on coordinates and momenta. 
PIMC calculations of the momentum distribution function and pair correlation functions 
for non-ideal plasma have been done. Comparison with classical Maxwell --  Boltzmann and quantum 
Fermi distribution shows the significant influence of the interparticle interaction on the high energy asymptotics
of the momentum distribution functions resulting in appearance of quantum 'tails'.  

\section{Acknowledgements}
We acknowledge stimulating discussions with Profs. A.N.~Starostin,  Yu.V.~Petrushevich, W.~Ebeling, M.~Bonitz, E.E.~Son, I.L.~Iosilevskii and V.I.~Man'ko.

%
%

\end{document}